\documentclass{article}
\usepackage[utf8]{inputenc}
\usepackage{xcolor,colortbl}
\usepackage{siunitx}
\usepackage{braket}
\usepackage{chemformula}
\usepackage{chemist} 
\usepackage{graphicx}
\usepackage{pdfpages}
\usepackage{hyperref}
\usepackage{authblk}
\usepackage{amsmath}
\usepackage{pdfpages}
\usepackage{braket}
\usepackage{siunitx}
\usepackage{multirow}
\usepackage{placeins}
\usepackage{amssymb}
\usepackage{geometry}
\usepackage{cite}    
\usepackage{ulem}
\usepackage{filecontents}

\newgeometry{vmargin={25mm}, hmargin={40mm,40mm}}   
\title{Associative detachment in anion-atom reactions involving a dipole-bound electron}
\author[1]{Saba Zia Hassan}
\author[1]{Jonas Tauch}
\author[2,3]{Milaim Kas}
\author[4]{Markus N\"otzold}
\author[1\dag]{Henry L\'opez Carrera}
\author[1,4]{Eric S. Endres}
\author[4]{Roland Wester}
\author[1*]{Matthias Weidem\"uller}
\affil[1]{Physikalisches Institut, Ruprecht-Karls-Universit\"at Heidelberg, 69120 Heidelberg, Germany}
\affil[2]{D\'epartement de Chimie, Facult\'e des Sciences, Universit\'e Libre de Bruxelles (ULB), 1050 Brusseles, Belgium}
\affil[3]{Deutsches Elektronen-Synchrotron (DESY), 22607 Hamburg, Germany}
\affil[4]{Institut f\"ur Ionenphysik und Angewandte Physik, Universit\"at Innsbruck, 6020 Innsbruck, Austria}
\affil[*]{Corresponding author: weidemueller@uni-heidelberg.de}
\affil[$\dag$]{Current address: Universidad de Las Fuerzas Armadas ESPE, 171103 Sangolqu\'i, Ecuador}

\begin{document}
\maketitle
\begin{abstract}
\noindent
Associative electronic detachment (AED) between anions and neutral atoms leads to the detachment of the anion's electron resulting in the formation of a neutral molecule. It plays a key role in chemical reaction networks, like the interstellar medium, the Earth’s ionosphere and biochemical processes. Here, a class of AED involving a closed-shell anion (OH$^-$) and alkali atoms (rubidium) is investigated by precisely controlling the fraction of  electronically  excited  rubidium. Reaction with the ground state atom gives rise to a stable intermediate complex with an electron solely bound via dipolar forces. The stability of the complex is governed by the subtle interplay of diabatic and adiabatic couplings into the autodetachment manifold. The measured rate coefficients are in good agreement with \textit{ab initio} calculations, revealing pronounced steric effects. For excited state rubidium, however, a lower reaction rate is observed, indicating dynamical stabilization processes suppressing the coupling into the autodetachment region. Our work provides a stringent test of \textit{ab initio} calculations on anion-neutral collisions and constitutes a generic, conceptual framework for understanding electronic state dependent dynamics in AEDs.
\end{abstract}
\newpage
\subsection*{Introduction}
Anions are ubiquitous in nature, from aqueous solution \cite{Lambert1968} and the earth's atmosphere \cite{Fehsenfeld1969} to astrochemical environments \cite{Bierbaum2011,Herbst1981,Millar2017}. They are reactive species, very sensitive to their environment, and often essential intermediates in important chemical events \cite{Simons2011}.
An important reactive process, distinguishing reactions involving anions from those involving cations or neutrals, is associative electronic detachment (AED), A + B$^{-}\rightarrow$ AB + e$^-$, which leads to the formation of a neutral molecule. The reaction is energetically allowed if the dissociation energy of AB is greater than the electron affinity of B. Given their universality, investigations of AED have led to profound understanding of phenomena in diverse chemical reaction networks. In the interstellar medium, AED is assumed to be one of the main destructive mechanisms of astrochemically relevant anions \cite{Eichelberger2007}. It serves as an intermediate step in the creation of complex molecules \cite{Millar2017,Wester2018}, contributes to the production of molecular hydrogen \cite{Kreckel2010} and the formation of interstellar water \cite{Plasil2017}. Furthermore, AED also plays a critical role in the formation of prebiotic molecules \cite{Phelps1968,Bierbaum2009,Bierbaum2019}. Extensive theoretical studies on the dynamics of AED exist for various systems, including halogen anions colliding with hydrogen \cite{Gauyacq1983}, the creation of hydroxyl molecules from O$^-$ and hydrogen \cite{Acharya1985}, the collisions of Li + H$^-$ \cite{Ek2018}, and the fundamental reaction H + H$^-$ \cite{Kreckel2010,Miller2011a}. In contrast, detailed experimental studies are limited to only few examples exploring reaction paths to the destruction of astrochemically relevant anions \cite{Fehsenfeld1969,Eichelberger2007,Kreckel2010,Miller2011a}.

Our work presents a detailed experimental investigation on the AED reaction dynamics between hydroxyl anions (OH$^-$) and a cloud of laser-cooled $^{85}$Rb atoms, in a hybrid atom-anion trap. The AED process in this system involving a closed shell anion and a single active electron atom is characterized by the emergence of an intermediate dipole-bound complex. Unlike valence-bound anions where the electron is characterized by dense, localized and multiply occupied orbitals, the excess electron in a dipole-bound anion lies in a very diffuse, singly-occupied orbital \cite{Desfrancois1996,Jordan2003,Lykke1984}. Historically, dipole-bound anions went from just being a theoretical curiosity to becoming identified as important species in various chemical processes, e.g electron capture in neutral molecules \cite{Skurski2002,Yuan2020}, zwitterion chemistry which plays an important role in amino acids \cite{Skurski2001b} and charge transfer processes \cite{Vila2002, Yandell2014}. In astrochemistry, dipole-bound anions have been invoked as important precursors to the formation of valence-bound anions \cite{Sommerfeld2005}, and as candidates for the explanation of diffuse interstellar bands \cite{Guthe2001,Simpson2021}. For our system, the intermediate dipole-bound anion exhibits a stable ground state and a short-lived excited state, resulting in a vastly different dynamics of the AED reaction. 

Over the last years, the study of controlled ion-neutral reactions has enabled insights into the collisional dynamics and the investigation of chemical phenomena at their most elementary level \cite{Smith2014,Cote2016,Puri2019,Paliwal2021}. However, most work focused on cationic and neutral systems, leaving out important collisions and reactions involving negative ions \cite{Dorfler2019}. Also, a comprehensive experimental study of the electronic state-dependence in controlled reactions is largely unexplored. 

In our work, making use of state-of-the-art techniques for trapping of ions and atoms \cite{Tomza2019,Puri2019}, we can precisely control the amount of excited rubidium, allowing us to explore the influence of the electronic state on the anion-neutral reaction dynamics.
The observed experimental results are compared to predictions of the Langevin classical capture model \cite{Langevin1905} and \textit{ab initio} calculations performed for the Rb-OH$^-$ system \cite{Kas2016,Byrd2013}. As we show, the Langevin model fails to explain the reaction dynamics for both the ground and excited state. In contrast, the \textit{ab initio} calculations, including steric effects, yield good quantitative agreement with the observed reaction rate coefficients for the ground state and provide a qualitative interpretation for the dynamics involving the excited state.

\begin{figure}[htp]
\centering{
\includegraphics[scale=0.75]{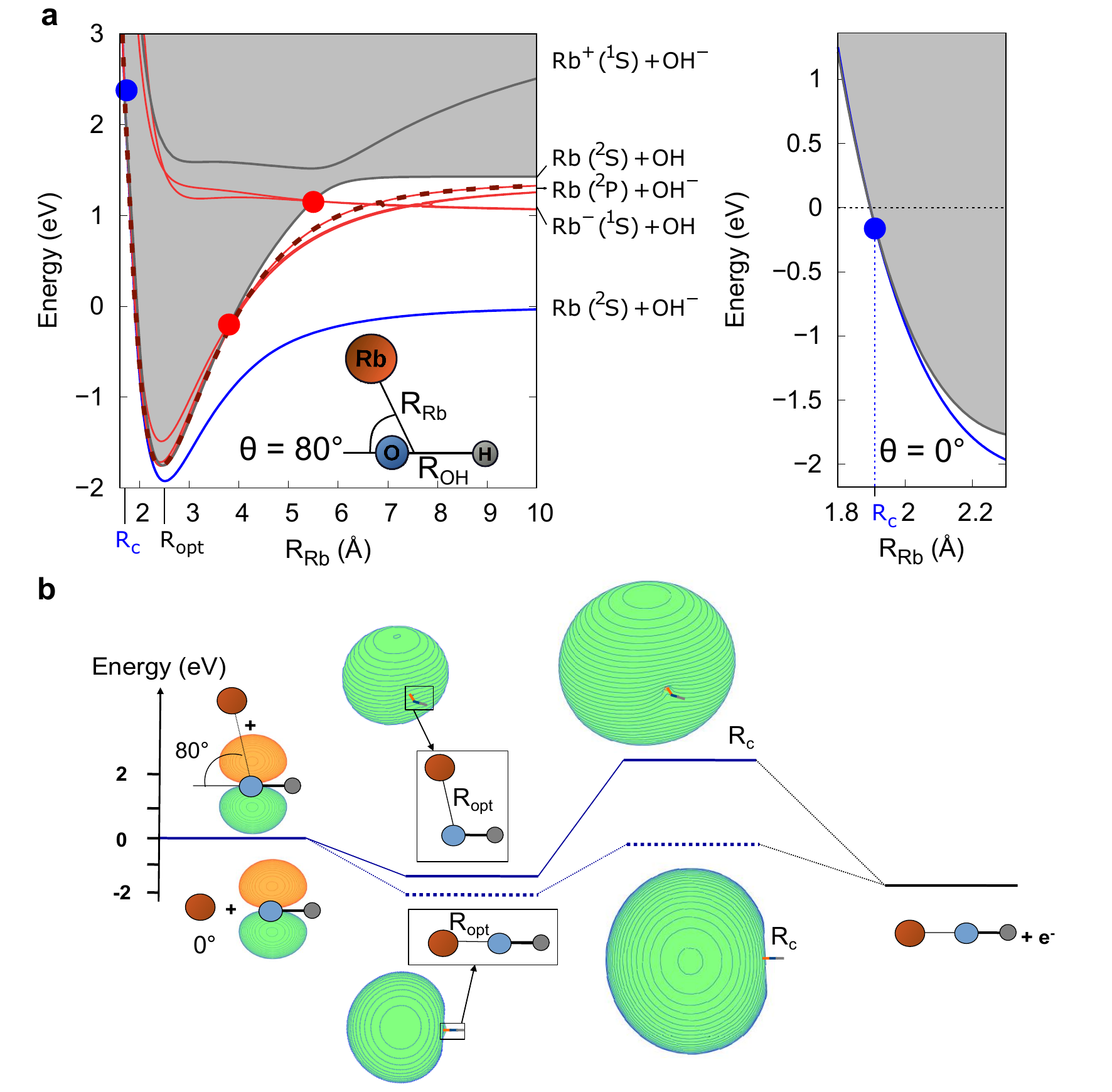}}
\caption{\textbf{Description of the Rb-OH$^-$ system.} \textbf{(a)} Potential energy surfaces as a function of the distance R$_\mathrm{Rb}$ between the Rb atom and the center of mass of O-H (left panel). The angle $\theta$ between the OH axis and the Rb atom is chosen to be $80^\circ$. The diabatic crossing between the excited state RbOH$^-$ complex (red curves) and its neutral counterpart RbOH (grey curve), is indicated by the two red dots. The crossing for the ground state RbOH$^-$ complex occurs at the inner part of the potentials (blue dot). R$_\text{opt}$ stands for the optimized R$_\text{Rb}$ distance (minimum of the interaction well) and R$_\text{c}$ corresponds to the distance at which the detachment occurs. The crossing in the repulsive region between the anionic and neutral PES for $\theta=0^\circ$ is shown in the right panel. \textbf{(b)} Reaction path for the AED reaction between ground state Rb and OH$^-$ shown for two different collisional angles: $\theta = 0^\circ$ (dashed blue line) and $\theta = 80^\circ$ (solid blue line). The zero corresponds to the energy of the Rb+OH$^-$ entrance channel. The orbital corresponding to the excess electron (highest occupied molecular orbital, HOMO) is shown.}
\label{fig:theo}
\end{figure}
\subsection*{Results}

\subsubsection*{Theoretical description of associative electronic detachment}
%
%
The only energetically accessible loss channel for Rb in the ground state is the AED reaction Rb($^2$S) + OH$^- \rightarrow $ RbOH + e$^-$, as theoretically investigated in \cite{Byrd2013,Kas2016}. For excited Rb, there are additional loss channels, of which the AED channel dominates (see Supplementary Note 4). In order to understand the underlying mechanism involved in the measured loss processes, we apply a modified Langevin capture model which takes into account that the AED is energetically allowed only for a finite range of angles of approach. For the ground state complex, the crossing between the anionic and neutral the potential energy surface (PES) occurs only at short range and not at long or intermediate internuclear distances due to the weak binding of the dipolar intermediate complex \cite{Byrd2013,Kas2016}. The potential energy surfaces (PESs) of the ground and low-lying electronic excited states of the anion Rb-OH$^-$ and neutral Rb-OH collisional complex have been calculated using \textit{ab initio} methods (for details see Methods and Supplementary Note 1).
A 1D cut of the PES at an exemplary collisional angle of $\theta=80^\circ$ is shown in the left panel of Figure \ref{fig:theo}a. The region for which the energy of the anion is larger than the energy of the neutral (when the anion PES intersects the neutral one) is defined as the autodetachment region (grey shaded area).

%
%
The ground state of the intermediate complex Rb-OH$^{-}$ (blue curve in Figure \ref{fig:theo}a) is stable against auto-detachment as its energy lies below the neutral one. It can be categorized as a dipole-bound state, despite its rather large detachment energy ($\approx 0.3$ eV) \cite{Kas2017}. The autodetachment region can only be reached for a limited angular space in the repulsive part of the PES (see the two cases of $\theta = 80^\circ$ and $0^\circ$ in Figure \ref{fig:theo}a, where the crossing is energetically inaccessible and accessible respectively). Thus, a much lower reaction rate than the upper bound given by the Langevin capture rate
is to be expected. 

More quantitatively, based on our \textit{ab initio} calculations (see \cite{Kas2016}), the reaction path for the AED reaction is shown in Figure \ref{fig:theo}b. The reaction starts with the Rb$+$OH$^-$ reactants, with the excess electron occupying a valence-bound $\pi$-orbital. The shape of the HOMO changes drastically when the reaction proceeds to the formation of the Rb-OH$^-$ intermediate complex exhibiting the typical halo shape of a dipolar bound complex.
The stability of a dipole-bound anion primarily depends on the dipole moment of the core. The dipole moment of the Rb-OH decreases with decreasing interatomic distance R$_\text{Rb}$ (as shown in the Supplementary Figure 2). Thus, a crossing with the energy of the neutral state can only occur in the repulsive inner region of the PES (see Figure \ref{fig:theo}a, right panel).

As the dipole moment increases with $\theta$, the anionic states become stabilized beyond a critical threshold value, which defines the angular space fraction $\rho$, in which the AED reaction occurs (see Methods for details). For $\theta < 20^\circ$ (dashed blue levels shown for $\theta = 0^\circ$), the energy crossing is found below the entrance channel threshold opening the AED channel (blue dot in Figure \ref{fig:theo}a right panel). For $\theta \gtrsim 20^\circ$ (solid blue levels shown in Figure \ref{fig:theo}b at $R_c$ for $\theta = 80^\circ$) the crossing becomes inaccessible.
In this case, $\rho$ is rather small, leading to a large deviation from the capture theory. The steep rising potential in the repulsive region, leads to a reaction probability highly sensitive to the \textit{ab initio} methods and, in particular, to the effective core potential used \cite{Byrd2013,Kas2016}. As shown in Table \ref{tab:results_all}, we find a reaction rate to be a factor of ten smaller than the Langevin rate employing the best available effective core potential for Rb. The calculation for the Langevin capture rate is described in Supplementary Note 3.

%
%
The electronic states of the dipole-bound intermediate complex, that are embedded into the autodetachement region, correlate to the excited entrance channel Rb($^2$P) + OH$^-$ (red curves in Figure \ref{fig:theo}a). Since all possible pathways lead to an energetically accessible crossing into the auto-detachment region, which ultimately leads to AED, the total loss rate is expected to be close to the capture rate.
Due to the highly diabatic nature of the potential energy surfaces, the dynamics of the AED reaction can be primarily described as following a single diabatic PES (red dashed curve in Figure \ref{fig:theo}a left panel). The excited state of the dipole-bound intermediate complex is short-lived and undergoes spontaneous autodetachment. 

However, for increasing collisional angles, the dipole moment increases and the excited state stabilizes similar to the ground state (see Supplementary Figure 2). In particular, above a critical collisional angle, here $\theta >150^\circ$, the crossing into the autodetachment region occurs at an energy higher than that of the entrance channel. This critical collisional angle defines the accessible angular space $\rho$, for AED with an excited state rubidium and is significantly larger than the ground state case.
The calculated loss rate is obtained using appropriate long-range interactions and features of the PES (see Methods). It is found to be close to the Langevin rate (see Table \ref{tab:results_all}), which is explained by the cancellation of additional long-range interaction terms and the reduced accessible angular space.

\subsubsection*{Measurement of reaction rate coefficients}
\begin{figure}[htp]
\centering{
\includegraphics[scale=0.75]{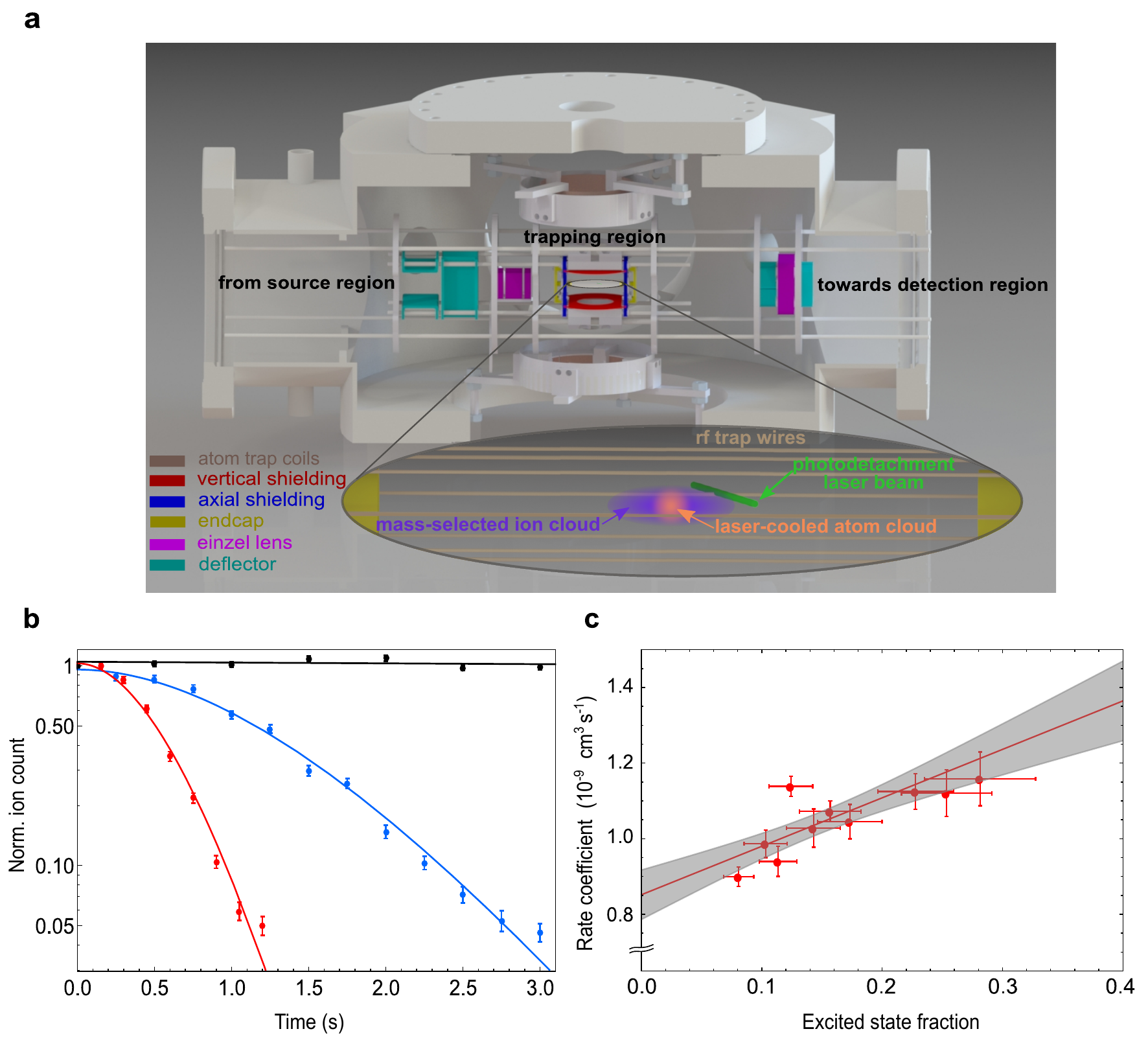}}
\caption{\textbf{Hybrid atom-ion system.} \textbf{(a)} The experimental hybrid atom ion trap system. The OH$^-$ ions (purple cloud) are created and loaded from the source chamber and trapped in an octupole radio-frequency (rf) wire trap. A far-threshold laser beam (green) is used to determine the ion density via photodetachment tomography. A laser-cooled cloud of ultracold Rb atoms (orange cloud) is overlapped with the ion cloud. As shown here, the spatial extent of the ion ensemble is significantly larger than that of the atomic cloud. After interaction with the laser or atoms, the time of flight of the ions are extracted onto the detector.  \textbf{(b)} The detected normalized OH$^-$ ion count after reaction with the laser-cooled rubidium atoms for excited state fractions of 0.10(2) and 0.28(5) (blue and red data points, respectively). The grey data points depict the ion losses without the presence of rubidium atoms. The ion losses are fitted by Equation \ref{eq:decay} (solid lines). The error bars represent the statistical errors. \textbf{(c)} Reaction rate coefficient as a function of the excited state fraction. The solid red line is a linear fit through the data points. The slope and intercept of the fit yields the reaction rate coefficient for the excited state and ground state rubidium interacting with OH$^-$, respectively. The grey shaded area represents the 1-$\sigma$ range of statistical uncertainty.}
\label{fig:exp}
\end{figure}

A mass-selected ensemble of hydroxyl anions OH$^-$ formed via electron attachment is loaded into an octupole radio-frequency wire trap, as schematically shown in Figure \ref{fig:exp}a and described in detail in \cite{Deiglmayr2012,haitrap2020}. Multipole ion traps feature a large field-free region in the radial direction, thus reducing radio-frequency heating \cite{Wester2009}. The hydroxyl ions occupy the vibrational ground state, as all higher vibrational states decay on a millisecond time scale. The kinetic temperature is set to $355(10)$\,K, via collisions with a pulse of helium buffer gas. The temperature of the ions is measured by mapping the ions' energy distribution to their time-of-flight (TOF) to the detector \cite{haitrap2020}. The ions' spatial distribution is mapped out by photodetachment tomography with a far-threshold laser \cite{Hlavenka2009}. 

Once the ions are trapped, they are overlapped with an ultracold cloud of rubidium atoms loaded into a dark spontaneous-force optical trap (dark-SPOT) configuration \cite{Ketterle1993}. In the dark-SPOT version of a magneto-optical trap, a part of the repumping laser beam is spatially blocked.
By changing the intensity of this laser beam, the fraction of atoms that are pumped back into the cooling cycle is controlled, thus providing control of the fraction of atoms in the excited (Rb($^2$P)) and ground state (Rb($^2$S)).
After a given reaction time, the number of ions remaining are extracted onto a detector, thus yielding a loss rate dependent on the fraction of electronically excited rubidium in the ensemble.

The ion losses for two different excited state fraction of the atom ensemble, are shown in Figure \ref{fig:exp}b. The evolution of the ion number in the ensemble, $N_I$, is expressed as: 
\begin{equation}
 N_I(t) = N^0_I \cdot \text{exp}\left( -\text{k} \int_0^t \Phi_{IA}(t') dt' \right) \cdot \text{exp}(-\text{k}_\text{bgd}t)
 \label{eq:decay}
\end{equation}
where $N^0_I$ is the initial ion number, k is the reaction rate coefficient, $\Phi_{IA}$ is the spatial overlap between the ion and atom cloud (see Methods) and k$_\text{bgd}$ is the background ion loss rate without the presence of atoms. The ion losses are fitted by Eq. \ref{eq:decay} yielding the reaction rate coefficient for the corresponding excited state fraction. By varying the amount of excited Rb in the atomic ensemble, a linear relationship between the reaction rate coefficients and excited state fractions is found as shown in Figure \ref{fig:exp}c. From the intercept of a linear fit through the data points, the reaction rate coefficient for Rb in the ground state is obtained as k$_\text{GS}$ = 8.5(7) $\times$ 10$^{-10}$\,cm$^3$s$^{-1}$ with the corresponding statistical uncertainty.
The slope yields the reaction rate coefficient for excited state as k$_\text{ES}$ = 2.1(4) $\times$ 10$^{-9}$\,cm$^3$s$^{-1}$. We estimate a systematic uncertainty of 40\% and 60\%, respectively, mainly due to the determination of the spatial overlap $\Phi_{IA}$ and parameters of the atom cloud.
\section*{Discussion}
The experimentally observed reaction rate coefficients are compared with the predictions of the classical capture theory and our \textit{ab initio} calculations (modified capture model) as shown in Table \ref{tab:results_all}.

%
%
The reaction rate for the ground state channel obtained from the experimental results shows the expected deviation from the Langevin prediction due to steric effects determining the stability of the intermediate dipolar complex, as previously outlined. 
The dipole moment drops below the critical threshold only for certain angles of approach for the anion-neutral collision, rendering a substantially reduced angular space where the autodetachment region is accessible. The experimental results confirm our understanding of the AED reaction dynamics, where for the first time, the dominant influence of a dipole-bound state as a critical reaction intermediate is revealed. 

In other alkali hydroxide anions, the autodetachment region is predicted to generally lie above the energy of the entrance channel \cite{Kas2017}. In such cases, depending on the energy gap between the anionic state and neutral state, diffuseness of the dipole-bound electron, and the reduced mass of the system, the AED rate is essentially determined by the presence of non-adiabatic coupling of the discrete states with the continuum \cite{Ek2018,Simons1998}. 

In general, alkali atoms when interacting with non-metal or halogen anionic species, will form complexes with large dipole moment due to the large difference in the electronegativity. These collisional complexes will most likely support dipole-bound excited states. Our results, thus provide a framework for probing the influence of stable dipole-bound states for future studies on anion-neutral reactions.

%
%
As shown in Table \ref{tab:results_all} for the excited state channel, the experimentally determined reaction rate deviates significantly from the capture model as well as the \textit{ab initio} calculations. This indicates the presence of additional stabilization mechanisms, like, e.g., the presence of longer lived metastable intermediate states in combination with non-adiabatic couplings, not accounted for in the current theoretical model.

In conclusion, we investigate the electronic quantum state-dependent AED rates in the Rb-OH$^-$ system. Through control on the electronic configuration of the rubidium atom, the electronic state of the intermediate dipolar complex is altered and its influence on the collisional detachment process is revealed. The intermediate complex is a dipole-bound anion which exhibits a stable ground state but short-lived excited states. A unifying feature of the reaction dynamics for both the ground and excited state channels is the accessible angular space which governs the probability of the reaction to occur. For the ground state Rb interacting with OH$^{-}$, the experimentally observed rate deviations from the capture model predictions are explained via the steric effects. Due to the high sensitivity of the reaction rate coefficient on the subtle details of the structure of the intermediate complex, our measurements provide a stringent test for different effective core potential models. However, for the excited state, the measured loss rate is significantly lower than the \textit{ab initio} and capture model predictions. A deeper theoretical investigation of the excited states of the dipole-bound Rb-OH$^-$ complex is needed to understand this discrepancy. Due to its similarity in the electronic structure of the intermediate complex to other alkali, alkali-earth hydroxides \cite{Kas2017}, and alkali with hydrated hydroxide \cite{Kas2019}, this work provides a general experimental framework to investigate state dependent alkali-anion reactions and opens new routes for the understanding of associative electronic detachment reactions. 

\newpage
\section*{Methods}
\begin{itemize}
\item \textbf{Determination of spatial overlap, $\Phi_{IA}(t)$ }
With an ion spatial density distribution constant in time (determined via ions' time-of-flight distribution), the evolution of spatial overlap between the ion and atom cloud is governed by the time-evolution of the atom cloud (loading behaviour) $n_A(t)=n^0_A\left(1-\text{exp}(-t/\tau_A)\right)$, where $n^0_A$ is the peak atom density and $\tau_A$ is the loading time. The volume of the atom cloud governs the boundary of the interaction region. The overlap $\Phi_{IA}(t)$ can be determined as:
\begin{equation}
 \Phi_{IA}(t)= \int \bar{n}_I(x,y,z) \cdot n_A(x,y,z,t) dx dy dz
\end{equation}
Here, $\bar{n}_I(x,y,z)$ is the unit-integral normalized OH$^-$ density and $n_A(x,y,z,t)$ is the time-dependent atom density distribution. The ion density distribution is determined via photodetachment tomography ~\cite{Trippel2006}. The potential landscape arising from the geometry of an octupole wire trap results in the ion density distribution radially proportional to  $r^{2n-2}$ (where $n=4$), and axially represented by a Gaussian profile. The atom cloud exhibits a Gaussian density profile determined via saturation absorption imaging \cite{Holtkemeier2018,Reinaudi2007}. The spatial extent of the ion cloud is much larger than that of the atomic cloud. The number of excited atoms are imaged by fluorescence imaging while the total number of atoms are imaged via saturation absorption imaging. The ratio of the two gives the excited state fraction of the atom cloud.

\item \textbf{Theoretical method}
The rate coefficients corresponding to the reaction involving Rb($^2\text{P}$) and OH$^-$ have been obtained using a modified capture model that includes features of the spin-orbit PES (see Supplementary Note 2), along with the following assumptions:
\begin{itemize}
\item Since the rubidium atom is present in its Rb($^2\text{P}_{3/2}$) fine state in the trap, the collision with OH$^-$ can either proceed following the 6$\text{E}_{1/2}$ or 5$\text{E}_{1/2}$ spin-orbit PES of the Rb-OH$^-$ complex. In the first case, the reaction proceeds on a potential that exhibits a barrier (see Supplementary Figure 4), in the second case the upper limit of the cross section is taken to be the capture cross section $\sigma_{Q}$ obtained with the following long-range potential: $V(R)=\frac{-\alpha}{2R^{4}}+\frac{-Q}{2R^{3}}+\epsilon(\frac{b}{R})^{2}$ where $\alpha$ and $Q$ are the static polarizability (870 a.u.) and quadrupole moment (26 a.u.) of Rb in its $^{2}\text{P}$ state.  
\item The detachment of the excess electron is assumed to be instantaneous when the autodetachment region is reached (sudden approximation). 
\item The transition probability between adiabatic potentials have been calculated using the Landau-Zener formula \cite{Zener1932}.
\item Electronic detachment can only be avoided if the collision takes place along the 2$\text{E}_{1/2}$ PES (first excited states of the Rb-OH$^-$ complex) within the angular space $\rho=\frac{1}{2}(1-cos(\theta_{max}))$ \cite{Kas2016}, where $\theta_{max}$ is the collision angle for which it crosses the neutral curve above the energy in the entrance channel. We found $\theta_{max} \approx 153^{\circ}$ (see Supplementary Figure 3). 
\item The  electronic to kinetic energy transfer reaction rate is expected to be very small. This reaction has therefore been neglected (see Supplementary Note 4).   
\end{itemize}\end{itemize}
The total loss cross section from the Rb($^{2}\text{P}_{3/2}$)+OH$^{-}(^{2}\Sigma^{+})$ entrance channel is given by a sum of the loss from the 5$\text{E}_{1/2}$ and $6\text{E}_{1/2}$ channels:
\begin{equation}
\sigma_{loss}(E_{c})=\frac{1}{2}\sigma^{6\text{E}_{1/2}}(E_{c})+\frac{1}{2}\sigma^{5\text{E}_{1/2}}(E_{c})
\end{equation}
The first and second term are obtained using the following expressions:
\begin{eqnarray}
\sigma^{6\text{E}_{1/2}}(E_{c})=\big(1-P_{NR}(E_{c})\,(1-\rho\,)\big)\,\sigma_{Q}(E_{c}) \\
\sigma^{5\text{E}_{1/2}}(E_{c})=\big(1-P_{NR}(E_{c})\,(1-\rho\,)\big)\,\sigma_{B}(E_{c})
\end{eqnarray}
where $(1-\rho)$ corresponds to the angular space where the crossing into the autodetachment region is avoided, $P_{NR}$ is the Landau-Zener probability to exit through the non-reactive channels Rb($^{2}\text{P}_{3/2}$) + OH$^{-}(^{1}\Sigma^{+})$ or Rb($^{2}\text{P}_{1/2}$) + OH$^{-}(^{1}\Sigma^{+})$  (thus 1-$P_{NR}$ is the probability to exit via the charge transfer channel), $\sigma_{Q}$ is the capture cross section and $\sigma_{B}=\pi R_{B}^{2}(1-U_{B}/E_{c})$ is a classical cross section. The latter is given in terms of the largest impact parameters $b_{max}$ for which the potential barrier is less than the collision energy $E_{c}$, $R_{B}=34 $ a.u. and $U_{B}=4.5\times 10^{-3}$ are the position and height of the potential barrier, respectively (see Supplementary Figure 4 ). The rate constant is then obtained by averaging over a Maxwell-Boltzmann distribution: 
\begin{equation}
k_{loss}=\sqrt{\frac{8}{\pi\mu(k_{b}T)^{3/2}}}\int \sigma_{loss}(E_{c})\,E_{c}\,e^{\frac{-E_{c}}{k_{b}T}}\,dE_{c}
\end{equation}  
where $\mu$ is the reduced mass of Rb-OH$^-$ system. \\

Owing to the highly diabatic nature of the PES ($i.e$ small Landau-Zener adiabatic transition probability, (see Supplementary Note 4)), the probability to exit through the charge transfer channel, $P_{CT}$, is very small, around $1.5\%$ for the relevant collision energies. Therefore $P_{NR}\approx 98.5\%$ and the entire dynamics is controlled by $\rho$. For $\rho = 1$, $\sigma_{loss} = \frac{1}{2}(\sigma_B+\sigma_Q)$ which leads to a capture case where all collisions that overcome the centrifugal and potential barrier, resulting in AED. For $\rho = 0$, the $2\text{E}_{1/2}$ state is stable against autodetachment (its energy lies below the neutral for all collision angles $\theta$). The total loss is then given by $\sigma_{loss} =\frac{1}{2}(1-P_{NR})(\sigma_B+\sigma_Q)$. Hence, associative detachment can only occur following the adiabatic states from the entrance channel for which the Landau-Zener probability is $(1-P_{NR})=1.5\times10^{-2}.$ With the factor of $0.5$, the total loss becomes $\sigma_{loss}= 7.5\times 10^{-3}(\sigma_B+\sigma_Q)$. 

\section*{Data Availability}
The data that support the findings of this study are available from the authors upon request.
\section*{Code Availability}
The codes and analysis files that support the findings of this study are available from the authors upon request.
\bibliographystyle{naturemag}
\bibliography{AD}

\begin{thebibliography}{10}
\expandafter\ifx\csname url\endcsname\relax
  \def\url#1{\texttt{#1}}\fi
\expandafter\ifx\csname urlprefix\endcsname\relax\def\urlprefix{URL }\fi
\providecommand{\bibinfo}[2]{#2}
\providecommand{\eprint}[2][]{\url{#2}}

\bibitem{Lambert1968}
\bibinfo{author}{Lambert, J.~L.}
\newblock \emph{\bibinfo{title}{{Trace inorganics in water}}},
  vol.~\bibinfo{volume}{73} of \emph{\bibinfo{series}{Adv. Chem. Ser.}},
  chap.~\bibinfo{chapter}{2}, \bibinfo{pages}{18--26}
  (\bibinfo{publisher}{American Chemical Society}, \bibinfo{year}{1968}).

\bibitem{Fehsenfeld1969}
\bibinfo{author}{Fehsenfeld, F.~C.}, \bibinfo{author}{Albritton, D.~L.},
  \bibinfo{author}{Burt, J.~A.} \& \bibinfo{author}{Schiff, H.~I.}
\newblock \bibinfo{title}{{Associative-detachment reactions of O$^-$ and
  O$_2^-$ by O$_2$($^1 \Delta_g$)}}.
\newblock \emph{\bibinfo{journal}{Can. J. Chem.}}
  \textbf{\bibinfo{volume}{47}}, \bibinfo{pages}{1793--1795}
  (\bibinfo{year}{1969}).

\bibitem{Bierbaum2011}
\bibinfo{author}{Bierbaum, V.~M.}
\newblock \bibinfo{title}{Anions in space and in the laboratory}.
\newblock \emph{\bibinfo{journal}{Proc. Int. Astron. Union}}
  \textbf{\bibinfo{volume}{7}}, \bibinfo{pages}{383–389}
  (\bibinfo{year}{2011}).

\bibitem{Herbst1981}
\bibinfo{author}{Herbst, E.}
\newblock \bibinfo{title}{{Can negative molecular ions be detected in dense
  interstellar clouds?}}
\newblock \emph{\bibinfo{journal}{Nature}} \textbf{\bibinfo{volume}{289}},
  \bibinfo{pages}{656--657} (\bibinfo{year}{1981}).

\bibitem{Millar2017}
\bibinfo{author}{Millar, T.~J.}, \bibinfo{author}{Walsh, C.} \&
  \bibinfo{author}{Field, T.~A.}
\newblock \bibinfo{title}{{Negative ions in space}}.
\newblock \emph{\bibinfo{journal}{Chem. Rev.}} \textbf{\bibinfo{volume}{117}},
  \bibinfo{pages}{1765--1795} (\bibinfo{year}{2017}).

\bibitem{Simons2011}
\bibinfo{author}{Simons, J.}
\newblock \bibinfo{title}{Theoretical study of negative molecular ions}.
\newblock \emph{\bibinfo{journal}{Annu. Rev. Phys. Chem.}}
  \textbf{\bibinfo{volume}{62}}, \bibinfo{pages}{107--128}
  (\bibinfo{year}{2011}).

\bibitem{Eichelberger2007}
\bibinfo{author}{Eichelberger, B.}, \bibinfo{author}{Snow, T.~P.},
  \bibinfo{author}{Barckholtz, C.} \& \bibinfo{author}{Bierbaum, V.~M.}
\newblock \bibinfo{title}{{Reactions of H, N, and O atoms with carbon chain
  anions of interstellar interest: An experimental study}}.
\newblock \emph{\bibinfo{journal}{ApJ}} \textbf{\bibinfo{volume}{667}},
  \bibinfo{pages}{1283--1289} (\bibinfo{year}{2007}).

\bibitem{Wester2018}
\bibinfo{author}{Jerosimić, S.~V.}, \bibinfo{author}{Gianturco, F.~A.} \&
  \bibinfo{author}{Wester, R.}
\newblock \bibinfo{title}{{Associative detachment (AD) paths for H and CN$^-$
  in the gas-phase: astrophysical implications}}.
\newblock \emph{\bibinfo{journal}{Phys. Chem. Chem. Phys.}}
  \textbf{\bibinfo{volume}{20}}, \bibinfo{pages}{5490--5500}
  (\bibinfo{year}{2018}).

\bibitem{Kreckel2010}
\bibinfo{author}{Kreckel, H.} \emph{et~al.}
\newblock \bibinfo{title}{{Experimental Results for H$_2$ Formation from H$^-$
  and H and implications for first star formation}}.
\newblock \emph{\bibinfo{journal}{Science}} \textbf{\bibinfo{volume}{329}},
  \bibinfo{pages}{69--71} (\bibinfo{year}{2010}).

\bibitem{Plasil2017}
\bibinfo{author}{Pla{\v{s}}il, R.} \emph{et~al.}
\newblock \bibinfo{title}{{Isotopic effects in the interaction of O$^-$ with
  D$_2$ and H$_2$ at low temperatures}}.
\newblock \emph{\bibinfo{journal}{Phys. Rev. A}} \textbf{\bibinfo{volume}{96}},
  \bibinfo{pages}{1--8} (\bibinfo{year}{2017}).

\bibitem{Phelps1968}
\bibinfo{author}{Moruzzi, J.~L.}, \bibinfo{author}{Ekin, J.~W.} \&
  \bibinfo{author}{Phelps, A.~V.}
\newblock \bibinfo{title}{{Electron production by associative detachment of
  O$^-$ ions with NO, CO, and H$_2$}}.
\newblock \emph{\bibinfo{journal}{J. Chem. Phys.}}
  \textbf{\bibinfo{volume}{48}}, \bibinfo{pages}{3070--3076}
  (\bibinfo{year}{1968}).

\bibitem{Bierbaum2009}
\bibinfo{author}{Snow, T.~P.} \emph{et~al.}
\newblock \bibinfo{title}{{Formation of gas-phase glycine and cyanoacetylene
  via associative detachment reactions}}.
\newblock \emph{\bibinfo{journal}{Astrobiology}} \textbf{\bibinfo{volume}{9}},
  \bibinfo{pages}{1001--1005} (\bibinfo{year}{2009}).

\bibitem{Bierbaum2019}
\bibinfo{author}{Nichols, C.~M.}, \bibinfo{author}{Wang, Z.-C.},
  \bibinfo{author}{Lineberger, W.~C.} \& \bibinfo{author}{Bierbaum, V.~M.}
\newblock \bibinfo{title}{{Gas-phase reactions of deprotonated nucleobases with
  H, N, and O atoms}}.
\newblock \emph{\bibinfo{journal}{J. Phys. Chem. Lett.}}
  \textbf{\bibinfo{volume}{10}}, \bibinfo{pages}{4863--4867}
  (\bibinfo{year}{2019}).

\bibitem{Gauyacq1983}
\bibinfo{author}{Gauyacq, J.~P.}
\newblock \bibinfo{title}{{Associative detachment and vibrational excitation in
  the e$^-$-HF system}}.
\newblock \emph{\bibinfo{journal}{J. Phys. B At. Mol. Phys.}}
  \textbf{\bibinfo{volume}{16}}, \bibinfo{pages}{4049--4058}
  (\bibinfo{year}{1983}).

\bibitem{Acharya1985}
\bibinfo{author}{Acharya, P.~K.}, \bibinfo{author}{Kendall, R.~A.} \&
  \bibinfo{author}{Simons, J.}
\newblock \bibinfo{title}{{Associative electron detachment: O$^-$ + H
  $\rightarrow$ OH + e$^-$}}.
\newblock \emph{\bibinfo{journal}{J. Chem. Phys.}}
  \textbf{\bibinfo{volume}{83}}, \bibinfo{pages}{3888--3893}
  (\bibinfo{year}{1985}).

\bibitem{Ek2018}
\bibinfo{author}{{\v{C}}{\'i}{\v{z}}ek, M.}, \bibinfo{author}{Dvo{\v{r}}{\'a}k,
  J.} \& \bibinfo{author}{Houfek, K.}
\newblock \bibinfo{title}{{Associative detachment in Li$+$ H$^-$ collisions}}.
\newblock \emph{\bibinfo{journal}{Eur. Phys. J. D}}
  \textbf{\bibinfo{volume}{72}}, \bibinfo{pages}{66} (\bibinfo{year}{2018}).

\bibitem{Miller2011a}
\bibinfo{author}{Miller, K.~A.} \emph{et~al.}
\newblock \bibinfo{title}{{Associative detachment of H${}^{\ensuremath{-}}$ + H
  $\ensuremath{\rightarrow}$ H${}_{2}$ + ${e}^{\ensuremath{-}}$}}.
\newblock \emph{\bibinfo{journal}{Phys. Rev. A}} \textbf{\bibinfo{volume}{84}},
  \bibinfo{pages}{052709} (\bibinfo{year}{2011}).

\bibitem{Desfrancois1996}
\bibinfo{author}{Desfran\ifmmode~\mbox{\c{c}}\else \c{c}\fi{}ois, C.}
  \emph{et~al.}
\newblock \bibinfo{title}{Prediction and observation of a new, ground state,
  dipole-bound dimer anion: The mixed water/ammonia system}.
\newblock \emph{\bibinfo{journal}{Phys. Rev. Lett.}}
  \textbf{\bibinfo{volume}{72}}, \bibinfo{pages}{48--51}
  (\bibinfo{year}{1994}).

\bibitem{Jordan2003}
\bibinfo{author}{Jordan, K.~D.} \& \bibinfo{author}{Wang, F.}
\newblock \bibinfo{title}{{Theory of dipole-bound anions}}.
\newblock \emph{\bibinfo{journal}{Annu. Rev. Phys. Chem.}}
  \textbf{\bibinfo{volume}{54}}, \bibinfo{pages}{367--396}
  (\bibinfo{year}{2003}).

\bibitem{Lykke1984}
\bibinfo{author}{Lykke, K.~R.}, \bibinfo{author}{Mead, R.~D.} \&
  \bibinfo{author}{Lineberger, W.~C.}
\newblock \bibinfo{title}{Observation of dipole-bound states of negative ions}.
\newblock \emph{\bibinfo{journal}{Phys. Rev. Lett.}}
  \textbf{\bibinfo{volume}{52}}, \bibinfo{pages}{2221--2224}
  (\bibinfo{year}{1984}).

\bibitem{Skurski2002}
\bibinfo{author}{Skurski, P.} \& \bibinfo{author}{Simons, J.}
\newblock \bibinfo{title}{{An excess electron bound to urea. III. The urea
  dimer as an electron trap}}.
\newblock \emph{\bibinfo{journal}{J. Chem. Phys.}}
  \textbf{\bibinfo{volume}{116}}, \bibinfo{pages}{6118--6125}
  (\bibinfo{year}{2002}).

\bibitem{Yuan2020}
\bibinfo{author}{Yuan, D.-F.} \emph{et~al.}
\newblock \bibinfo{title}{Observation of a $\ensuremath{\pi}$-type dipole-bound
  state in molecular anions}.
\newblock \emph{\bibinfo{journal}{Phys. Rev. Lett.}}
  \textbf{\bibinfo{volume}{125}}, \bibinfo{pages}{073003}
  (\bibinfo{year}{2020}).

\bibitem{Skurski2001b}
\bibinfo{author}{Skurski, P.} \& \bibinfo{author}{Simons, J.}
\newblock \bibinfo{title}{{An excess electron bound to urea. I. Canonical and
  zwitterionic tautomers}}.
\newblock \emph{\bibinfo{journal}{J. Chem. Phys.}}
  \textbf{\bibinfo{volume}{115}}, \bibinfo{pages}{8373--8380}
  (\bibinfo{year}{2001}).

\bibitem{Vila2002}
\bibinfo{author}{Vila, F.~D.} \& \bibinfo{author}{Jordan, K.~D.}
\newblock \bibinfo{title}{{Theoretical study of the dipole-bound excited states
  of I$^-$(H$_2$O)$_4$}}.
\newblock \emph{\bibinfo{journal}{J. Phys. Chem. A}}
  \textbf{\bibinfo{volume}{106}}, \bibinfo{pages}{1391--1397}
  (\bibinfo{year}{2002}).

\bibitem{Yandell2014}
\bibinfo{author}{Yandell, M.~A.}, \bibinfo{author}{King, S.~B.} \&
  \bibinfo{author}{Neumark, D.~M.}
\newblock \bibinfo{title}{Decay dynamics of nascent acetonitrile and
  nitromethane dipole-bound anions produced by intracluster charge-transfer}.
\newblock \emph{\bibinfo{journal}{J. Chem. Phys.}}
  \textbf{\bibinfo{volume}{140}}, \bibinfo{pages}{184317}
  (\bibinfo{year}{2014}).

\bibitem{Sommerfeld2005}
\bibinfo{author}{Sommerfeld, T.}
\newblock \bibinfo{title}{{Dipole-bound states as doorways in (dissociative)
  electron attachment}}.
\newblock \emph{\bibinfo{journal}{J. Phys. Conf. Ser.}}
  \textbf{\bibinfo{volume}{4}}, \bibinfo{pages}{245--250}
  (\bibinfo{year}{2005}).

\bibitem{Guthe2001}
\bibinfo{author}{Guthe, F.}, \bibinfo{author}{Tulej, M.},
  \bibinfo{author}{Pachkov, M.~V.} \& \bibinfo{author}{Maier, J.~P.}
\newblock \bibinfo{title}{{Photodetachment spectrum of I-C$_3$H$_2^-$: The role
  of dipole bound states for electron attachment in interstellar clouds}}.
\newblock \emph{\bibinfo{journal}{ApJ}} \textbf{\bibinfo{volume}{555}},
  \bibinfo{pages}{466--471} (\bibinfo{year}{2001}).

\bibitem{Simpson2021}
\bibinfo{author}{Simpson, M.} \emph{et~al.}
\newblock \bibinfo{title}{{Influence of a Supercritical Electric Dipole Moment
  on the Photodetachment of ${\mathrm{C}}_{3}{\mathrm{N}}^{\ensuremath{-}}$}}.
\newblock \emph{\bibinfo{journal}{Phys. Rev. Lett.}}
  \textbf{\bibinfo{volume}{127}}, \bibinfo{pages}{043001}
  (\bibinfo{year}{2021}).

\bibitem{Smith2014}
\bibinfo{author}{Smith, W.~W.} \emph{et~al.}
\newblock \bibinfo{title}{Experiments with an ion-neutral hybrid trap: cold
  charge-exchange collisions}.
\newblock \emph{\bibinfo{journal}{Appl. Phys. B}}
  \textbf{\bibinfo{volume}{114}}, \bibinfo{pages}{75--80}
  (\bibinfo{year}{2014}).

\bibitem{Cote2016}
\bibinfo{author}{{C{\^{o}}t{\'{e}}, R.}}
\newblock \bibinfo{title}{{Chapter two - Ultracold hybrid atom–ion systems}}.
\newblock vol.~\bibinfo{volume}{65} of \emph{\bibinfo{series}{Adv. At. Mol.
  Opt. Phys.}}, \bibinfo{pages}{67--126} (\bibinfo{publisher}{Academic Press},
  \bibinfo{year}{2016}).

\bibitem{Puri2019}
\bibinfo{author}{Puri, P.} \emph{et~al.}
\newblock \bibinfo{title}{Reaction blockading in a reaction between an excited
  atom and a charged molecule at low collision energy}.
\newblock \emph{\bibinfo{journal}{Nat. Chem.}} \textbf{\bibinfo{volume}{11}},
  \bibinfo{pages}{615--621} (\bibinfo{year}{2019}).

\bibitem{Paliwal2021}
\bibinfo{author}{Paliwal, P.} \emph{et~al.}
\newblock \bibinfo{title}{Determining the nature of quantum resonances by
  probing elastic and reactive scattering in cold collisions}.
\newblock \emph{\bibinfo{journal}{Nat. Chem.}} \textbf{\bibinfo{volume}{13}},
  \bibinfo{pages}{94--98} (\bibinfo{year}{2021}).

\bibitem{Dorfler2019}
\bibinfo{author}{D{\"o}rfler, A.~D.} \emph{et~al.}
\newblock \bibinfo{title}{{Long-range versus short-range effects in cold
  molecular ion-neutral collisions}}.
\newblock \emph{\bibinfo{journal}{Nat. Commun.}} \textbf{\bibinfo{volume}{10}},
  \bibinfo{pages}{5429} (\bibinfo{year}{2019}).

\bibitem{Tomza2019}
\bibinfo{author}{Tomza, M.} \emph{et~al.}
\newblock \bibinfo{title}{Cold hybrid ion-atom systems}.
\newblock \emph{\bibinfo{journal}{Rev. Mod. Phys.}}
  \textbf{\bibinfo{volume}{91}}, \bibinfo{pages}{035001}
  (\bibinfo{year}{2019}).

\bibitem{Langevin1905}
\bibinfo{author}{Langevin, P.}
\newblock \bibinfo{title}{{A fundamental formula of kinetic theory.}}
\newblock \emph{\bibinfo{journal}{Ann. Chim. Phys.}}
  \textbf{\bibinfo{volume}{5}}, \bibinfo{pages}{245--288}
  (\bibinfo{year}{1905}).

\bibitem{Kas2016}
\bibinfo{author}{Kas, M.}, \bibinfo{author}{Loreau, J.},
  \bibinfo{author}{Li\'{e}vin, J.} \& \bibinfo{author}{Vaeck, N.}
\newblock \bibinfo{title}{{Ab initio study of reactive collisions between
  Rb($^2$S) or Rb($^2$P) and OH$^-$(1\Sigma$^+$)}}.
\newblock \emph{\bibinfo{journal}{J. Chem. Phys.}}
  \textbf{\bibinfo{volume}{144}}, \bibinfo{pages}{204306}
  (\bibinfo{year}{2016}).

\bibitem{Byrd2013}
\bibinfo{author}{Byrd, J.~N.}, \bibinfo{author}{Michels, H.~H.},
  \bibinfo{author}{Montgomery, J.~A.} \& \bibinfo{author}{C{\^{o}}t{\'{e}}, R.}
\newblock \bibinfo{title}{{Associative detachment of rubidium hydroxide}}.
\newblock \emph{\bibinfo{journal}{Phys. Rev. A - At. Mol. Opt. Phys.}}
  \textbf{\bibinfo{volume}{88}}, \bibinfo{pages}{032710}
  (\bibinfo{year}{2013}).

\bibitem{Kas2017}
\bibinfo{author}{Kas, M.}, \bibinfo{author}{Loreau, J.},
  \bibinfo{author}{Li{\'{e}}vin, J.} \& \bibinfo{author}{Vaeck, N.}
\newblock \bibinfo{title}{{Ab initio study of the neutral and anionic alkali
  and alkaline earth hydroxides: Electronic structure and prospects for
  sympathetic cooling of OH$^-$}}.
\newblock \emph{\bibinfo{journal}{J. Chem. Phys.}}
  \textbf{\bibinfo{volume}{146}}, \bibinfo{pages}{194309}
  (\bibinfo{year}{2017}).

\bibitem{Deiglmayr2012}
\bibinfo{author}{Deiglmayr, J.}, \bibinfo{author}{G{\"{o}}ritz, A.},
  \bibinfo{author}{Best, T.}, \bibinfo{author}{Weidem{\"{u}}ller, M.} \&
  \bibinfo{author}{Wester, R.}
\newblock \bibinfo{title}{{Reactive collisions of trapped anions with ultracold
  atoms}}.
\newblock \emph{\bibinfo{journal}{Phys. Rev. A - At. Mol. Opt. Phys.}}
  \textbf{\bibinfo{volume}{86}}, \bibinfo{pages}{043438}
  (\bibinfo{year}{2012}).

\bibitem{haitrap2020}
\bibinfo{author}{Nötzold, M.} \emph{et~al.}
\newblock \bibinfo{title}{Thermometry in a multipole ion trap}.
\newblock \emph{\bibinfo{journal}{Appl. Sci.}} \textbf{\bibinfo{volume}{10}},
  \bibinfo{pages}{5264} (\bibinfo{year}{2020}).

\bibitem{Wester2009}
\bibinfo{author}{Wester, R.}
\newblock \bibinfo{title}{{Radiofrequency multipole traps: Tools for
  spectroscopy and dynamics of cold molecular ions}}.
\newblock \emph{\bibinfo{journal}{J. Phys. B At. Mol. Opt. Phys.}}
  \textbf{\bibinfo{volume}{42}}, \bibinfo{pages}{154001}
  (\bibinfo{year}{2009}).

\bibitem{Hlavenka2009}
\bibinfo{author}{Hlavenka, P.} \emph{et~al.}
\newblock \bibinfo{title}{{Absolute photodetachment cross section measurements
  of the O$^-$ and OH$^-$ anion}}.
\newblock \emph{\bibinfo{journal}{J. Chem. Phys.}}
  \textbf{\bibinfo{volume}{130}}, \bibinfo{pages}{061105}
  (\bibinfo{year}{2009}).

\bibitem{Ketterle1993}
\bibinfo{author}{Ketterle, W.}, \bibinfo{author}{Davis, K.~B.},
  \bibinfo{author}{Joffe, M.~A.}, \bibinfo{author}{Martin, A.} \&
  \bibinfo{author}{Pritchard, D.~E.}
\newblock \bibinfo{title}{{High densities of cold atoms in a dark
  spontaneous-force optical trap}}.
\newblock \emph{\bibinfo{journal}{Phys. Rev. Lett.}}
  \textbf{\bibinfo{volume}{70}}, \bibinfo{pages}{2253--2256}
  (\bibinfo{year}{1993}).

\bibitem{Simons1998}
\bibinfo{author}{Simons, J.}
\newblock \bibinfo{title}{{Semiquantum expressions for electronically
  nonadiabatic electron ejection rates}}.
\newblock \emph{\bibinfo{journal}{J. Phys. Chem. A}}
  \textbf{\bibinfo{volume}{102}}, \bibinfo{pages}{6035--6042}
  (\bibinfo{year}{1998}).

\bibitem{Kas2019}
\bibinfo{author}{Kas, M.}, \bibinfo{author}{Loreau, J.},
  \bibinfo{author}{Liévin, J.} \& \bibinfo{author}{Vaeck, N.}
\newblock \bibinfo{title}{{Reactivity of hydrated hydroxide anion clusters with
  H and Rb: An ab initio study}}.
\newblock \emph{\bibinfo{journal}{J. Phys. Chem. A}}
  \textbf{\bibinfo{volume}{123}}, \bibinfo{pages}{8893--8906}
  (\bibinfo{year}{2019}).

\bibitem{Trippel2006}
\bibinfo{author}{Trippel, S.} \emph{et~al.}
\newblock \bibinfo{title}{{Photodetachment of cold OH$^-$ in a multipole ion
  trap}}.
\newblock \emph{\bibinfo{journal}{Phys. Rev. Lett.}}
  \textbf{\bibinfo{volume}{97}}, \bibinfo{pages}{193003}
  (\bibinfo{year}{2006}).

\bibitem{Holtkemeier2018}
\bibinfo{author}{H{\"{o}}ltkemeier, B.}, \bibinfo{author}{Gl{\"{a}}ssel, J.},
  \bibinfo{author}{L{\'{o}}pez-Carrera, H.} \&
  \bibinfo{author}{Weidem{\"{u}}ller, M.}
\newblock \bibinfo{title}{{A dense gas of laser-cooled atoms for hybrid
  atom-ion trapping}}.
\newblock \emph{\bibinfo{journal}{Appl. Phys. B Lasers Opt.}}
  \textbf{\bibinfo{volume}{123}}, \bibinfo{pages}{51} (\bibinfo{year}{2017}).

\bibitem{Reinaudi2007}
\bibinfo{author}{Reinaudi, G.}, \bibinfo{author}{Lahaye, T.},
  \bibinfo{author}{Wang, Z.} \& \bibinfo{author}{Gu{\'{e}}ry-Odelin, D.}
\newblock \bibinfo{title}{{Strong saturation absorption imaging of dense clouds
  of ultracold atoms}}.
\newblock \emph{\bibinfo{journal}{Opt. Lett.}} \textbf{\bibinfo{volume}{32}},
  \bibinfo{pages}{3143--3145} (\bibinfo{year}{2007}).

\bibitem{Zener1932}
\bibinfo{author}{Zener, C.}
\newblock \bibinfo{title}{{Non-adiabatic crossing of energy levels}}.
\newblock \emph{\bibinfo{journal}{Proc. R. Soc. A Math. Phys. Eng. Sci.}}
  \textbf{\bibinfo{volume}{137}}, \bibinfo{pages}{696--702}
  (\bibinfo{year}{1932}).

\end{thebibliography}


\begin{thebibliography}{10}
\expandafter\ifx\csname url\endcsname\relax
  \def\url#1{\texttt{#1}}\fi
\expandafter\ifx\csname urlprefix\endcsname\relax\def\urlprefix{URL }\fi
\providecommand{\bibinfo}[2]{#2}
\providecommand{\eprint}[2][]{\url{#2}}

\bibitem{Berning2000}
\bibinfo{author}{Berning, A.}, \bibinfo{author}{Schweizer, M.},
  \bibinfo{author}{Werner, H.-J.}, \bibinfo{author}{Knowles, P.~J.} \&
  \bibinfo{author}{Palmieri, P.}
\newblock \bibinfo{title}{{Spin-orbit matrix elements for internally contracted
  multireference configuration interaction wavefunctions}}.
\newblock \emph{\bibinfo{journal}{Mol. Phys.}} \textbf{\bibinfo{volume}{98}},
  \bibinfo{pages}{1823--1833} (\bibinfo{year}{2000}).

\bibitem{Knowles1992}
\bibinfo{author}{Knowles, P.~J.} \& \bibinfo{author}{Werner, H.-J.}
\newblock \bibinfo{title}{{Internally contracted multiconfiguration-reference
  configuration interaction calculations for excited states}}.
\newblock \emph{\bibinfo{journal}{Theor. Chim. Acta}}
  \textbf{\bibinfo{volume}{84}}, \bibinfo{pages}{95--103}
  (\bibinfo{year}{1992}).

\bibitem{Langhoff1974}
\bibinfo{author}{Langhoff, S.~R.} \& \bibinfo{author}{Davidson, E.~R.}
\newblock \bibinfo{title}{{Configuration interaction calculations on the
  nitrogen molecule}}.
\newblock \emph{\bibinfo{journal}{Int. J. Quantum Chem.}}
  \textbf{\bibinfo{volume}{8}}, \bibinfo{pages}{61--72} (\bibinfo{year}{1974}).

\bibitem{Lim2005}
\bibinfo{author}{Lim, I.~S.}, \bibinfo{author}{Schwerdtfeger, P.},
  \bibinfo{author}{Metz, B.} \& \bibinfo{author}{Stoll, H.}
\newblock \bibinfo{title}{{All-electron and relativistic pseudopotential
  studies for the group 1 element polarizabilities from K to element 119}}.
\newblock \emph{\bibinfo{journal}{J. Chem. Phys.}}
  \textbf{\bibinfo{volume}{122}}, \bibinfo{pages}{104103}
  (\bibinfo{year}{2005}).

\bibitem{Dunning1988}
\bibinfo{author}{Dunning, T.~H.}
\newblock \bibinfo{title}{{Gaussian basis sets for use in correlated molecular
  calculations. I. The atoms boron through neon and hydrogen}}.
\newblock \emph{\bibinfo{journal}{J. Chem. Phys.}}
  \textbf{\bibinfo{volume}{90}}, \bibinfo{pages}{1007} (\bibinfo{year}{1989}).

\bibitem{Sansonetti2006}
\bibinfo{author}{Sansonetti, J.~E.}
\newblock \bibinfo{title}{{Wavelengths, Transition Probabilities, and Energy
  Levels for the Spectra of Rubidium (Rb I through Rb XXXVII)}}.
\newblock \emph{\bibinfo{journal}{J. Phys. Chem. Ref. Data}}
  \textbf{\bibinfo{volume}{35}}, \bibinfo{pages}{301} (\bibinfo{year}{2006}).

\bibitem{Mies1973}
\bibinfo{author}{Mies, F.~H.}
\newblock \bibinfo{title}{{Molecular theory of atomic collisions:
  Fine-structure transitions}}.
\newblock \emph{\bibinfo{journal}{Phys. Rev. A}} \textbf{\bibinfo{volume}{7}},
  \bibinfo{pages}{942--957} (\bibinfo{year}{1973}).

\bibitem{Langevin1905}
\bibinfo{author}{Langevin, P.}
\newblock \bibinfo{title}{{A fundamental formula of kinetic theory.}}
\newblock \emph{\bibinfo{journal}{Ann. Chim. Phys.}}
  \textbf{\bibinfo{volume}{5}}, \bibinfo{pages}{245} (\bibinfo{year}{1905}).

\bibitem{vogt1954}
\bibinfo{author}{Vogt, E.} \& \bibinfo{author}{Wannier, G.~H.}
\newblock \bibinfo{title}{Scattering of ions by polarization forces}.
\newblock \emph{\bibinfo{journal}{Phys. Rev.}} \textbf{\bibinfo{volume}{95}},
  \bibinfo{pages}{1190--1198} (\bibinfo{year}{1954}).

\bibitem{Mitroy2010}
\bibinfo{author}{Mitroy, J.}, \bibinfo{author}{Safronova, M.~S.} \&
  \bibinfo{author}{Clark, C.~W.}
\newblock \bibinfo{title}{Theory and applications of atomic and ionic
  polarizabilities}.
\newblock \emph{\bibinfo{journal}{J. Phys. B: At. Mol. Opt. Phys.}}
  \textbf{\bibinfo{volume}{43}}, \bibinfo{pages}{202001}
  (\bibinfo{year}{2010}).

\bibitem{Kas2016}
\bibinfo{author}{Kas, M.}, \bibinfo{author}{Loreau, J.},
  \bibinfo{author}{Li\'{e}vin, J.} \& \bibinfo{author}{Vaeck, N.}
\newblock \bibinfo{title}{{Ab initio study of reactive collisions between
  Rb($^2$S) or Rb($^2$P) and OH$^-$(1\Sigma$^+$)}}.
\newblock \emph{\bibinfo{journal}{J. Chem. Phys.}}
  \textbf{\bibinfo{volume}{144}}, \bibinfo{pages}{204306}
  (\bibinfo{year}{2016}).

\bibitem{Byrd2013}
\bibinfo{author}{Byrd, J.~N.}, \bibinfo{author}{Michels, H.~H.},
  \bibinfo{author}{Montgomery, J.~A.} \& \bibinfo{author}{C{\^{o}}t{\'{e}}, R.}
\newblock \bibinfo{title}{{Associative detachment of rubidium hydroxide}}.
\newblock \emph{\bibinfo{journal}{Phys. Rev. A - At. Mol. Opt. Phys.}}
  \textbf{\bibinfo{volume}{88}}, \bibinfo{pages}{032710}
  (\bibinfo{year}{2013}).

\bibitem{Nikitin1999}
\bibinfo{author}{Nikitin, E.~E.}
\newblock \bibinfo{title}{{Nonadiabatic transitions： what we learned from old
  masters and how much we owe them}}.
\newblock \emph{\bibinfo{journal}{Annu. Rev. Phys. Chem. 1999.}}
  \textbf{\bibinfo{volume}{50}}, \bibinfo{pages}{1} (\bibinfo{year}{1999}).

\bibitem{Desouter-Lecomte1977}
\bibinfo{author}{Desouter-Lecomte, M.} \& \bibinfo{author}{Lorquet, J.~C.}
\newblock \bibinfo{title}{{Nonadiabatic interactions in unimolecular decay. II.
  Simplified formalism}}.
\newblock \emph{\bibinfo{journal}{J. Chem. Phys.}}
  \textbf{\bibinfo{volume}{66}}, \bibinfo{pages}{4006--4017}
  (\bibinfo{year}{1977}).

\end{thebibliography}
\section*{Acknowledgments}
The authors would like to acknowledge Dr. Bastian H\"oltkemeier for his role in the realization of the experimental setup. This work is supported by the Austrian Science Fund (FWF) through Project No. I3159-N36 and Deutsche Forschungsgemeinschaft (DFG) under Project No. WE/2661/14-1. S.Z.H. acknowledges the support from IMPRS-QD fellowship and HGSFP. M.K. is grateful to the BMBF project MeSoX (Project No. 05K19GUE) for financial support.
\section*{Author Contributions}
R.W. and M.W. conceived the project, S.Z.H., J.T. and H.L.C. performed experimental design and implementation, S.Z.H., M.N. and E.E. analysed the experimental results, M.K. performed the theoretical investigations, S.Z.H. and M.K. wrote the manuscript. All authors contributed to the editing and revision of the manuscript.
\section*{Competing Interests}
The authors declare no competing interests.
\newpage
\section*{Tables}
\begin{table}[htb]
\caption{Comparison of results from the classical capture theory predictions (Langevin) and the \textit{ab initio} calculations (modified capture model), to the experimentally obtained reaction rate coefficients (at $355$\,K). k$_\text{GS}$ and k$_\text{ES}$ are the ground state and excited state reaction rate coefficients, respectively.}
 \begin{center}
  \begin{tabular}{ | l | c | c | } 
   \hline
       & k$_\text{GS}$ (10$^{-9}$\,cm$^3$s$^{-1}$)   & k$_\text{ES}$ (10$^{-9}$\,cm$^3$s$^{-1}$)\\
                \hline
   Langevin                 & 4.3     & 7.2     \\
   \textit{Ab initio} calculations  & 0.42    & 7.5     \\
   Experiment               & 0.85(7) & 2.1(4)  \\
   \hline
  \end{tabular} 
 \end{center}

 \label{tab:results_all}
\end{table}
\end{document}


\begin{center}
\section*{\LARGE{Supplementary Information}}\end{center}
{\let\newpage\relax\maketitle}
\newpage
\setcounter{section}{0}
\section*{Supplementary Note 1}
\section*{\textit{Ab-initio} potential energy surfaces}
\label{theory_ESPES}
\label{theory_abinitio}
Multi-configuration approaches have been used in order to obtain the excited potential energy surfaces that correlate to the Rb(P$_{3/2}$)+OH$^-$ entrance channel. We have accounted for the spin-orbit coupling using the state interacting method as implemented in the \begin{small}MOLPRO\end{small} \cite{Berning2000} program using the spin-orbit operator from the MDF effective core potential. A total of 6 interacting anionic states and 2 neutral states have been taken into account. A set of state averaged orbitals was obtained by performing multi-configuration calculation on selected configurations \cite{Knowles1992}. These orbitals were then used to perform configuration interaction using the internally contracted scheme (ic-MRCI) implemented in the \begin{small}MOLPRO\end{small} program, including the Davidson correction (labelled +Q) \cite{Langhoff1974}.\\
The ic-MRCI+Q has been applied to a set of manually selected configurations, in the C$_{s}$ point group. The selected configurations have been obtained from a separate SA-CASSCF calculation on the four $\text{A}'$ and two $\text{A}''$ low lying electronic states of the RbOH$^{-}$ molecular complex, taken in its optimized geometry (linear geometry). The active space covers 9$a'$ and 3$a''$ molecular orbitals (MO).\\
Singly excited configurations were found to be dominant contributors in the CI expansion. This is expected from electronic configurations consideration. The ground-state electronic configuration is $1a'^{2}2a'^{2}3a'^{2}1a''^{2}4a'^{2}5a'^{2}6a'^{2}7a'^{2}2a''^{2}8a'^{1}$, where the $1a'$, $2a'$ and $3a'$ MOs correspond to the $1s_{O}$, $4s_{Rb}$ and $2s_{O}$ atomic orbital (AO), respectively, the $4a'$, $1a''$, $5a'$ and $6a'$, $7a'$, $2a''$ MOs are mainly formed by the $4p_{Rb}$ and $2p_{O}$ AOs, respectively, and the $8a'$ MO is mainly formed by the $5s_{Rb}$ AO and corresponds to the highest occupied MO (HOMO). In further discussion we will omit the $1a'^{2}2a'^{2}3a'^{2}1a''^{2}4a'^{2}5a'^{2}6a'^{2}$ inner shell, which will be labelled with brackets [], and only focus on the valence MOs. The first unoccupied MOs, $9a'$, $10a'$ and $3a''$ are mainly formed by the $5p_{Rb}$ AOs, the ground-state electronic configuration becomes $[]7a'^{2}2a''^{2}8a'^{1}9a'^{0}10a'^{0}3a''^{0}$. The first four $\text{A}'$ and two $\text{A}''$ low lying excited states of the RbOH$^{-}$ molecular complex correlate to the following channels: Rb$^{-}(^{1}\text{S})$+OH$(^{2}\Pi)$, Rb$(^{2}\text{S})$+OH$^{-}(^{1}\Sigma^{+})$ and Rb$(^{2}\text{P})$+OH$^{-}(^{1}\Sigma^{+})$. Their main electronic configuration correspond to single excitation relative to the ground-state configuration within the valence $7a'2a''8a'9a'10a'3a''$ MOs. The different electronic states along with their main electronic configurations are given in Supplementary Table \ref{CONF}. The dissociation channels at which they diabatically correlate are also depicted. It should be noted that these electronic excited states correspond to shape resonances of the RbOH+$e^{-}$ collisional system. 

\begin{table}[h]
\renewcommand{\arraystretch}{1.5}
\begin{tabular}{|c|c|l|}
\hline
\textbf{Electronic states} & \textbf{Main electronic config.} & \textbf{Dissociation Height} \\
\hline
X$\,^{2}$A' & $[]7a'^{2}2a''^{2}8a'^{1}$ & Rb($^{2}\text{S}$)+OH$^{-}$($^{1}\Sigma^{+}$) \\ \hline
2$\,^{2}$A'$\oplus$1$\,^{2}$A'' & $[]7a'^{2}2a''^{2}8a'^{0}9a'^{1}$ $\oplus$ $[]7a'^{2}2a''^{2}8a'^{0}3a''^{1}$ & \multirow{2}{*}{Rb($^{2}\text{P}$)+OH$^{-}$($^{1}\Sigma^{+}$)}  \\
3$\,^{2}$A' & $[]7a'^{2}2a''^{2}8a'^{0}10a'^{1}$ &  \\ \hline
4$\,^{2}$A'$\oplus$2$\,^{2}$A'' & $[]7a'^{1}2a''^{2}8a'^{2}$ $\oplus$ $[]7a'^{2}2a''^{1}8a'^{2}$ & Rb$^{-}$($^{1}\text{S}$)+OH($^{2}\Pi$)\\ \hline 
\end{tabular}
\caption{The main electronic configurations and dissociation channels of different electronic states are presented.}
\label{CONF}
\end{table}

A particularly interesting feature of the Rb-OH$^-$ collisional complex is the drastic change in the binding nature of the excess electron along the reaction path. The excess electron is first localized on the hydroxyl anion, which is a typical closed shell valence-bound anion, characterized by a compact density, large detachment energies ($\approx 1.8$ eV) and important electron correlation effects. When the reaction proceed, the Rb atom disturbs the electron cloud, the excess electron becomes more diffuse and less strongly bound where it is primary bound via dipole-charge interaction. \\
One of the usual difficulties one faces while using standard bound-state quantum chemistry approaches to calculated metastable states embedded in a continuum is that the wave function usually undergoes variational collapses to a neutral + free electron state where the excess electron occupies a very diffuse orbital. Various stabilization schemes can be used to tackle this problem. Here we found out that constraining the reference MCSCF wave function to single excited configurations avoids the convergence problems and allows us to obtain the resonance states even when using diffuse basis set. The potential energy surfaces (PESs) shown in the main text have been obtained with this manually selected MCSCF/ic-MRCI approach using the MDF ECP for Rb along with its $spdfg$ companion basis set \cite{Lim2005} supplemented by a set of $5s4p3d2fg$ even tempered diffuse functions and the AVQZ basis set for the O and H atom \cite{Dunning1988}.\\ Using group theory consideration the four $^{1}\text{A}'$ + two $^{1}\text{A}''$ electronic states result in six interacting $\text{E}_{1/2}$ spin-orbit (SO) states. The results are shown in Supplementary Figure \ref{PES_SO} for $\theta=80^{\circ}$. The two insets show the behaviour near the potential well and at dissociation, lower and upper left panels, respectively. The PES of the neutral RbOH ground-state is shown in black line. Our calculation correctly reproduces the SO splitting of the $^{2}\text{P}_{3/2}$ and $^{2}\text{P}_{1/2}$ state of Rb, with a calculated value of 26.5 \si{\milli\electronvolt} which is to be compared to experimental value 29.3 \si{\milli\electronvolt} \cite{Sansonetti2006}.
\begin{figure}[h!]
\includegraphics[scale=0.82]{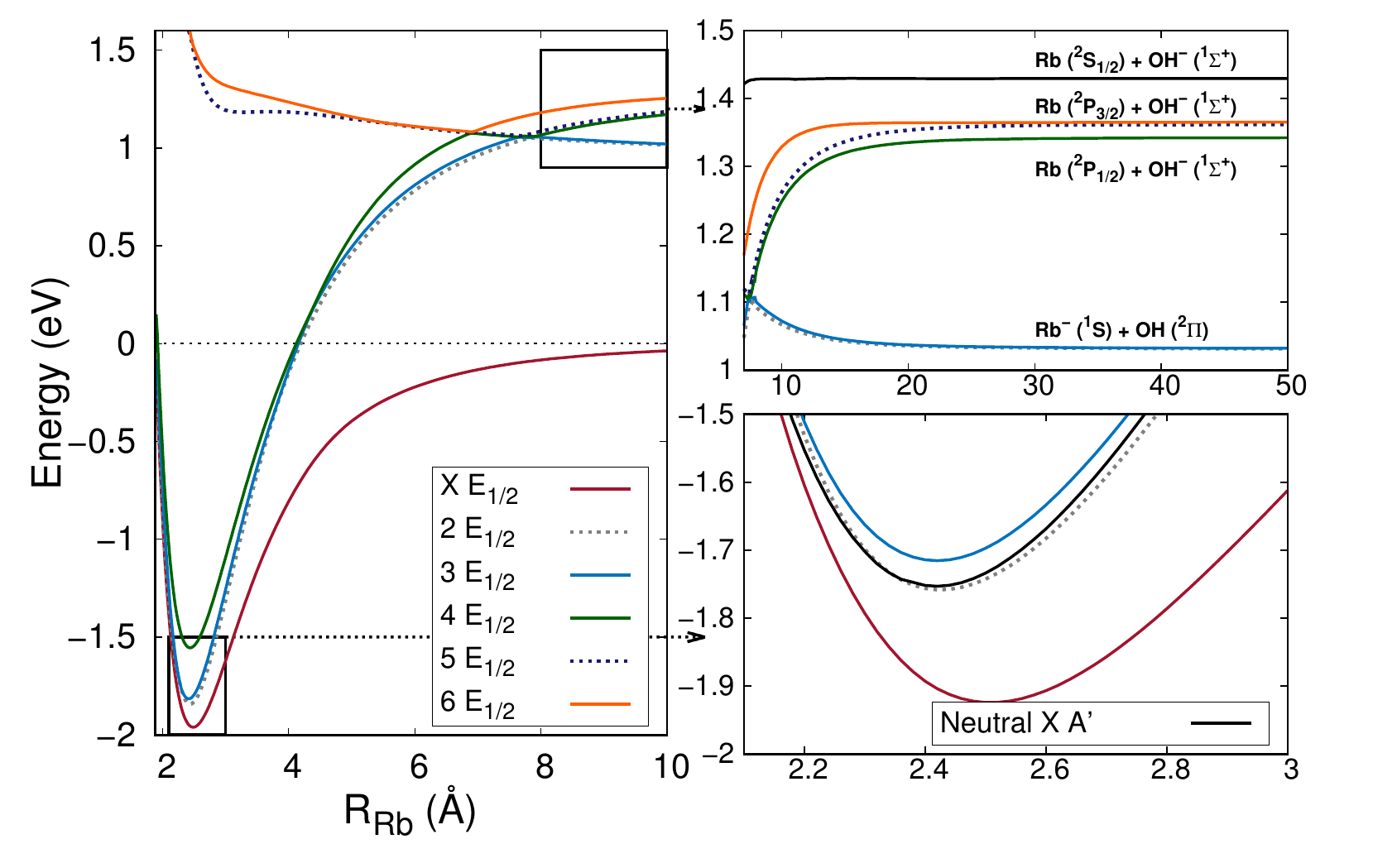}
\caption{Left panel: PESs of the low-lying excited states of the Rb-OH$^-$ molecular complex for $\theta=80^\circ$, including SO couplings. This leads to 6 interacting $\text{E}_{1/2}$ states. The red curve and other solid colored lines correspond to the adiabatic potential of the ground- and excited-states of the Rb-OH$^-$ collision complex, respectively. The dashed lines correspond to the $\text{A}"$ states in the non-relativistic picture. The upper right panel show the PESs of the excited states at dissociation where the relevant channels are depicted. The bottom right panel shows the PESs of the 3 lowest anionic states (red, dashed gray and blue curve) around the potential well alongside the neutral potential (black curve).}
\label{PES_SO}
\end{figure}
\section*{Supplementary Note 2}
\section*{Features of the potential energy surface}
\label{theory_desPES}
\setcounter{section}{2}
\subsection{Potential well region}
Our calculations show that the excited states of the RbOH$^{-}$ complex, taken in its equilibrium geometry ($\theta = 0^\circ$) are auto-detaching states, in other words, only the ground state has a positive vertical detachment energy (VDE). However, for larger values of $\theta$ dipole moment of the RbOH core increases, leading to a stabilization of the anion. This can be seen in Supplementary Figure \ref{fig:dipole}, where the the Hartree-Fock dipole moment of the RbOH core is depicted as a function of R$_\text{Rb}$ for $\theta=0^\circ$ (black) and as a function of $\theta$ at optimized R$_\text{Rb}$ distance (blue). 
\begin{figure}
    \centering
    \includegraphics[scale=0.4]{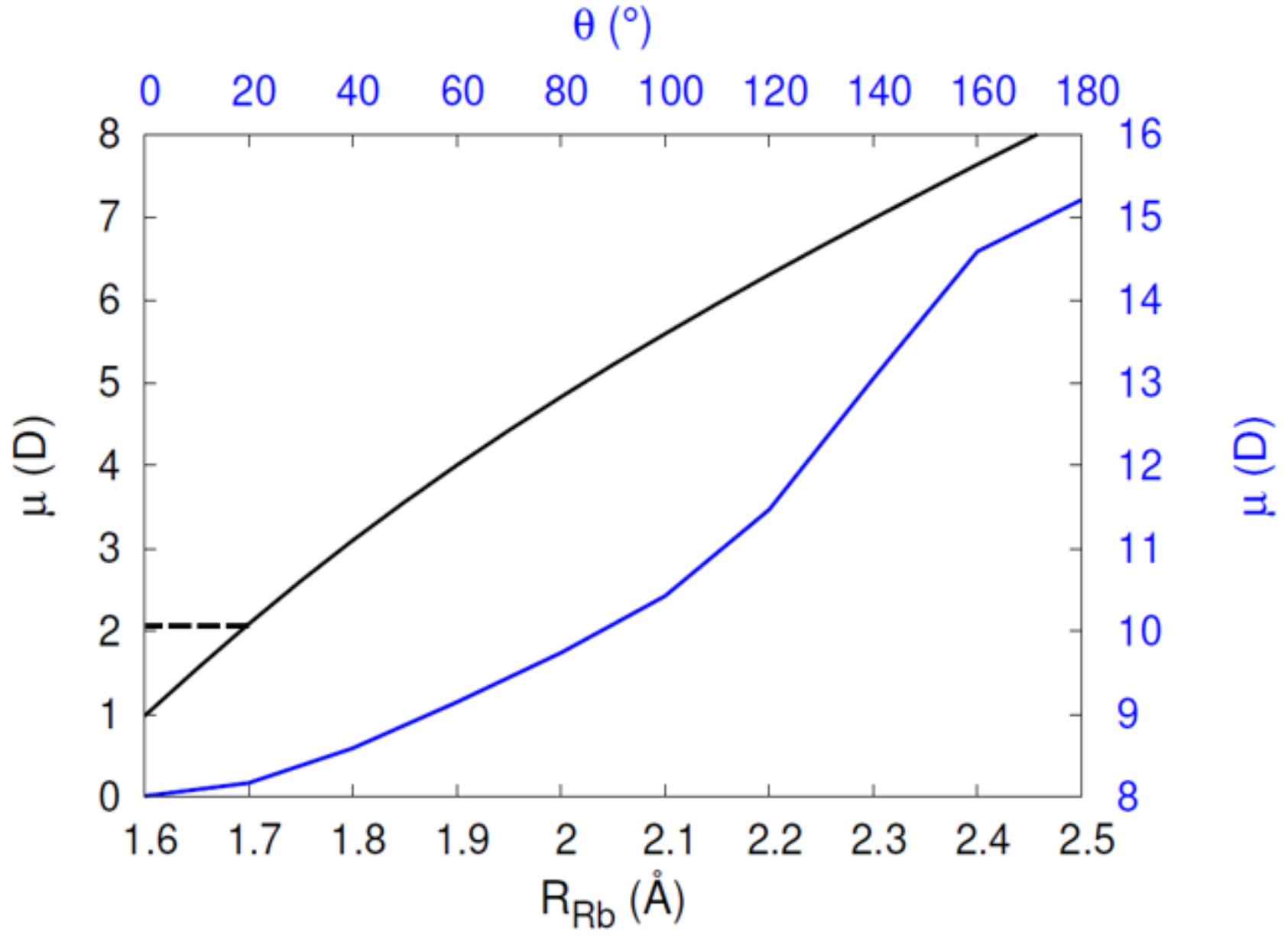}
    \caption{Hartree-Fock calculation of the dipole moment of RbOH. The black curve shows the dipole moment $\mu$ as a function of internuclear distance R$_\text{Rb}$ for $\theta=0^\circ$.  The blue curve shows the dipole moment as a function of collisional angle $\theta$ for an optimized internuclear distance. The critical value of the dipole moment ensuring a stable intermediate dipolar complex is indicated by the dashed line.}
    \label{fig:dipole}
\end{figure}
As a consequence, the VDE increases for increasing $\theta$. This trend can be seen in Supplementary Figure \ref{2ApN} where the crossing point between the anionic first excited-state $2\text{E}_{1/2}$ PES and the neutral PES is marked by a dot. In particular, for $\theta>153^\circ$ the crossing with the neutral PES occurs above the Rb($^{2}\text{P}_{3/2}$)+OH$^{-}(^{1}\Sigma^{+})$ entrance channel energy. The later is taken as $E_c+T(\mu,J)$, where $E_c$ and $T(\mu,J)$ are the collision energy and internal energy (vibration and rotation) of OH$^-$ at $T=355$K, respectively. This characterizes an accessible angular space where the auto-detachment from the 2$\text{E}_{1/2}$ is avoided.
\begin{figure}[h]
\centering
\includegraphics[scale=0.65]{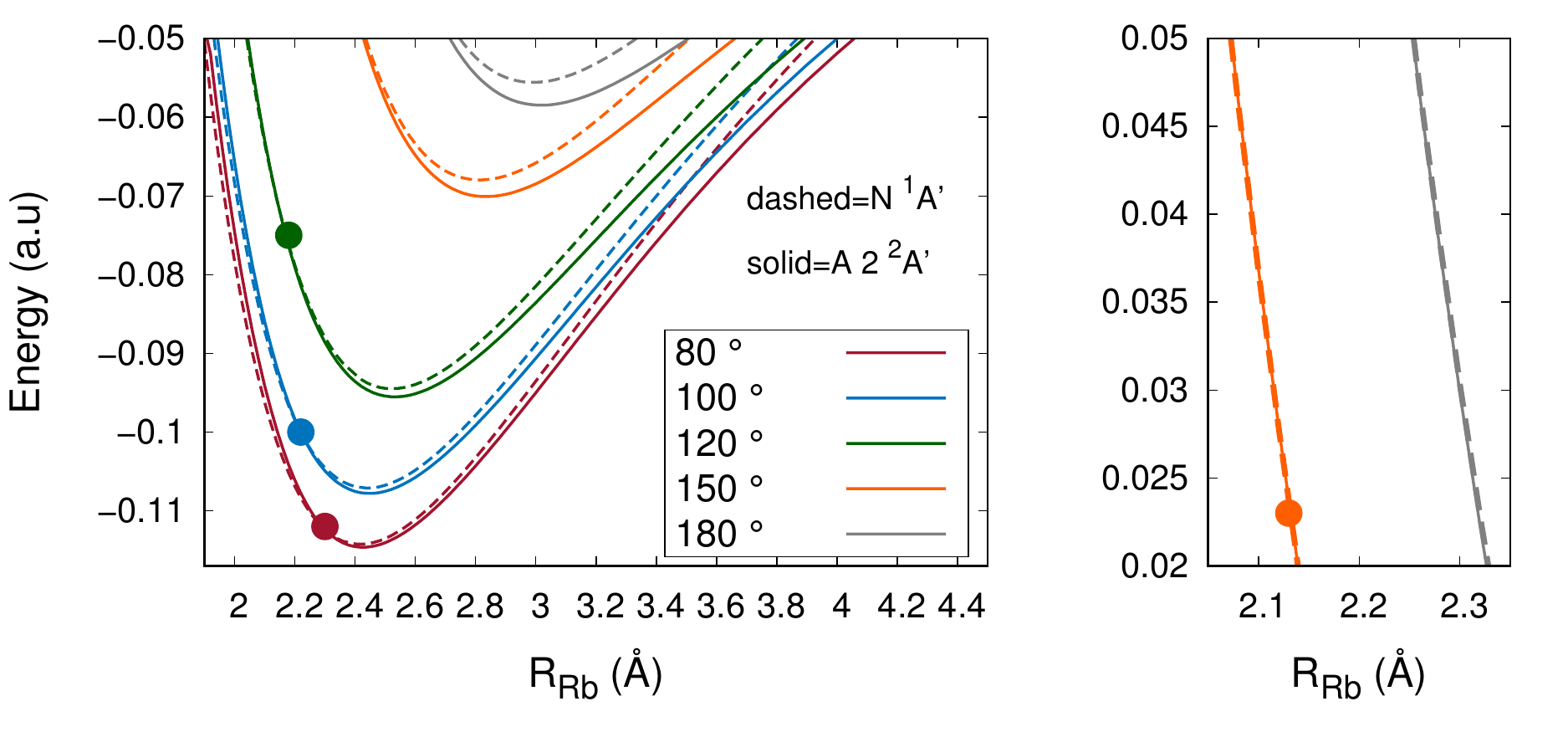}
\caption{PESs for the first excited state of the anion RbOH$^-$ ($2\,^{2}\text{A}'$ or $2\,\text{E}_{1/2}$, solid lines) and the ground state of the neutral RbOH molecular complex (dashed lines), for various collision angle $\theta$. Note that the colors corresponds to various values of $\theta$. The right panel shows a zoom-in of the PESs in the repulsive region. The zero energy is taken as the energy of the ground state entrance channel Rb($^{2}\text{S}$)+OH$^{-}$.}
\label{2ApN}
\end{figure}
\subsection{Long and intermediate range}
Although it is difficult to see in the Supplementary Figure \ref{PES_SO}, the 6$\text{E}_{1/2}$ state is repulsive at long range and attractive at intermediate distance, leading to a potential barrier around 14\,\si{\angstrom}. The barrier can be seen in Supplementary Figure \ref{barrier} where the PESs for the 6$\text{E}_{1/2}$ state for various values of $\theta$ have been plotted. The barrier height is almost independent of $\theta$ and is about $4.5\times10^{-3}$\,\si{\electronvolt} high. This repulsive state can be related to the classical repulsive potential of a charge-quadrupole interaction \cite{Mies1973}. We suspect the presence of a higher lying state, with a repulsive potential at intermediate and short range, that interacts with the 6$\text{E}_{1/2}$ state leading to an avoided crossing. We found this state to be mainly described by doubly excited electronic configurations, hence it corresponds to a Feshbach resonance of the RbOH+$e^{-}$ collisional complex. Our approach based on manually selected singly excited configurations is not adequate to describe this additional state. Unfortunately, due to the convergence problems pointed out above, adding even the minimum number of doubly excited configurations needed to describe this additional state results in variational collapses of the wave function. Therefore, specific methods would be needed.\\   

\begin{figure}[h]
\centering
\includegraphics[scale=0.8]{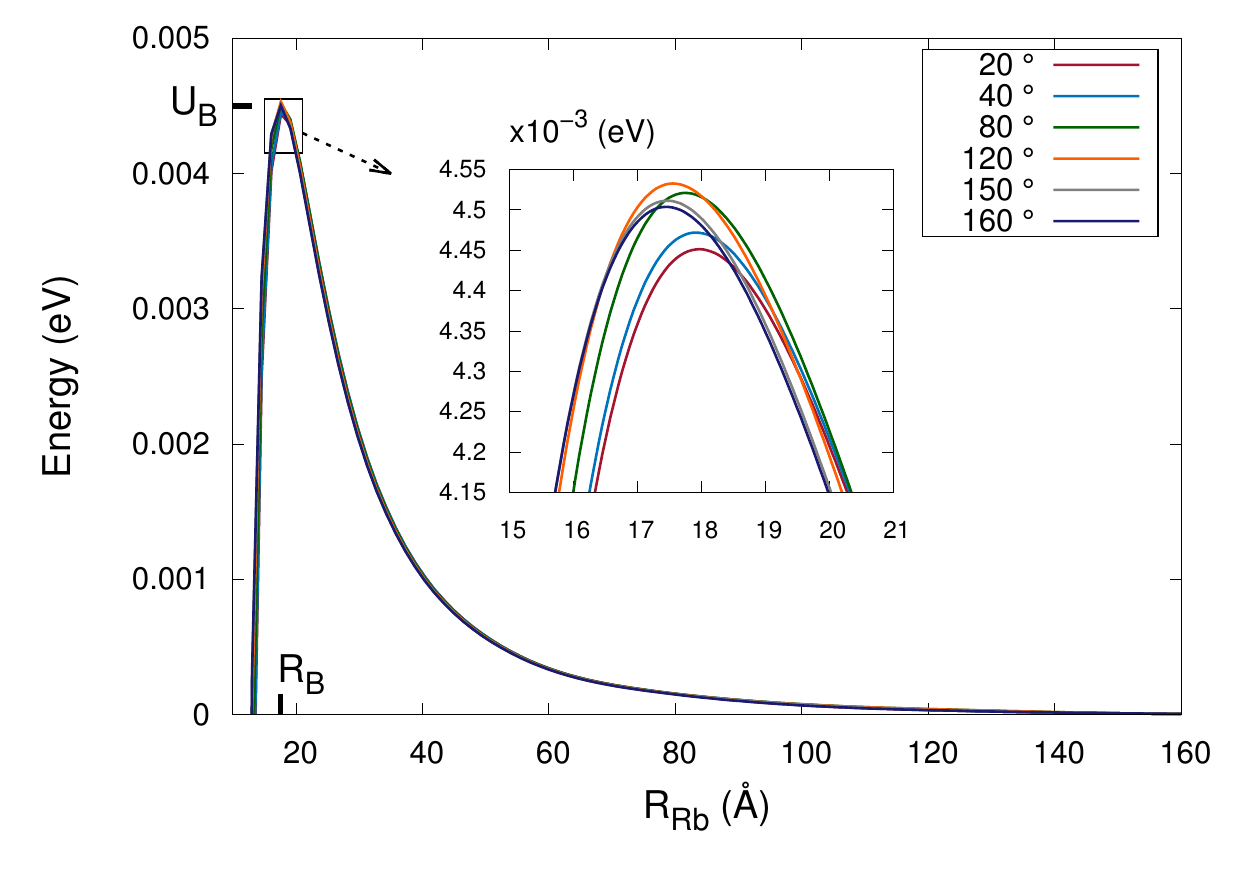}
\caption{Potential barrier along the 6$\text{E}_{1/2}$ state for various values of $\theta$. The distance R$_\text{B}$ is the intermolecular distance R$_\text{Rb}$ at the potential's maximum value U$_\text{B}$.}
\label{barrier}
\end{figure}

\section*{Supplementary Note 3}
\section*{Determination of Langevin reaction rate}
\label{langevin}
In the event of an atom-ion collision, the upper limit to collisional rate constants is given by the classical-mechanics based, Langevin capture model \cite{Langevin1905}. The long range interaction potential V($r$) is dominated by the interaction between charge-induced dipole of the atom with the ion's charge such that:
 \begin{align}
V_\text{int}(r)=-\frac{C_4}{r^4}
\end{align}
where $r$ is the interparticle separation and $C_4=\frac{\alpha e^2}{(4\pi\epsilon_0)^2}$\cite{vogt1954}. Here $\alpha$ is the scalar polarizability of the neutral, $e$ is the electron charge and $\epsilon_0$ is the permittivity of free space.\\
By including the centrifugal potential into the effective potential of the collision complex, the Langevin reaction rate constant $k_\text{L}$ can be derived as :
\begin{align}
k_\text{L}=2\pi\sqrt{\frac{C_4}{\mu}}=\frac{e}{2\epsilon_0}\sqrt{\frac{\alpha}{\mu}}
\end{align}
where all parameters are in S.I. units.
\\ The scalar polarizabilities and the calculated reaction rate constants for excited and ground-state Rb are summarized as follows:
\begin{table}[h!]
\centering
\begin{tabular}{|c|c|c|}
\hline 
& $\alpha$ \text{ (a.u.)} \cite{Mitroy2010} & k$_\text{L}$ ($ 10^{-9}$ \text{cm}$^3$\text{s}$^{-1}$) \\ 
\hline 
Rb ($^2$P) & 870  & 7.2  \\ 
\hline 
Rb ($^2$S) & 318.6  & 4.3 \\ 
\hline
\end{tabular} 
\label{tableLangevin} 
\end{table}\\
\textbf{Note:} Conversion factor used for $\alpha$ (from atomic units to S.I. units) \cite{Mitroy2010},
$1 \text{ a.u.}=1.648773 \times 10^{-41}$ \si{\coulomb\m\squared\per\volt}.
\\ 
The modified Langevin model that has been used to calculate the loss rate from the ground and excited state channel averages over the range of angles of approach for which the reaction is exoergic. There are some implicit assumptions in this model, which require some additional discussion

\begin{enumerate}
        \item The present model assumes that no angle of approach is preferred. Steering effects, i.e. reorientation of the OH$^-$ molecule due to intermolecular forces, start to play a role at low temperatures and would lead to an increase in the likelihood of a "head-on" collision of Rb-O-H, thus increasing the AED rate towards the Langevin rate. Although we do not expect these effects to be dominant in the temperature regime of our experiment, they might partially explain the underestimation of the measured rate coefficient by the \emph{ab initio} calculation.
        \item The population of rotational levels in OH$^-$ may not follow a thermal distribution due to collisional cooling with the ultracold buffer gas. However, it has been shown that a lower rotational temperature would actually decrease the measured AED rate \cite{Kas2016}.   
        \item As suggested in \cite{Byrd2013}, the presence of vibrational excited OH$^{-}$ anions will strongly increase the AED loss. Using the dipole moment of OH$^-$, we find that the lifetime of $v=1$ state, before radiative decay is on the order of couple of milliseconds, which in comparison to the complete timescale of the experiment ($\sim 10$ seconds), makes the fraction of ions in the excited vibrational states negligible, as mentioned in the main text.
        
\end{enumerate}


\section*{Supplementary Note 4}
\section*{Excited-state loss channels}
\label{esd}
The interaction of hydroxyl anion with excited Rb ($^2$P) opens the following loss channels:
\begin{itemize}
\label{ES_loss}
\item \textbf{Associative electronic detachment (AED):}
\begin{center}
Rb($^2$P) + OH$^-$ $\rightarrow$ RbOH + e$^-$ 
\end{center}
\item \textbf{Electronic to kinetic energy transfer:}
\begin{center}
Rb($^2$P) + OH$^-$ $\rightarrow$ Rb + OH$^-$ + E$_{kin}$
\end{center}
\item \textbf{Charge-exchange reaction:}
\begin{center}
Rb($^2$P) + OH$^- \rightarrow $ Rb$^-$ + OH
\end{center}
\end{itemize}
\setcounter{section}{4}
\setcounter{subsection}{0}
\subsection{Electronic to kinetic energy transfer}

In order to estimate the probability to exit through the electronic to kinetic energy transfer Rb($^{2}\text{P}_{3/2}$)+OH$^{-} \rightarrow$ Rb($^{2}\text{S}$)+OH$^{-}$ channel, we have calculated the non-adiabatic coupling matrix elements (NACME) between the X$\text{E}_{1/2}$ and $2\text{E}_{1/2}$ states, which link to entrance and exit channels. This has been done using the finite difference approach, implemented in \begin{small}MOLPRO\end{small}. The results can be seen in Supplementary Figure \ref{NACME}. 
\begin{figure}[h]
\begin{center}
\includegraphics[scale=0.2]{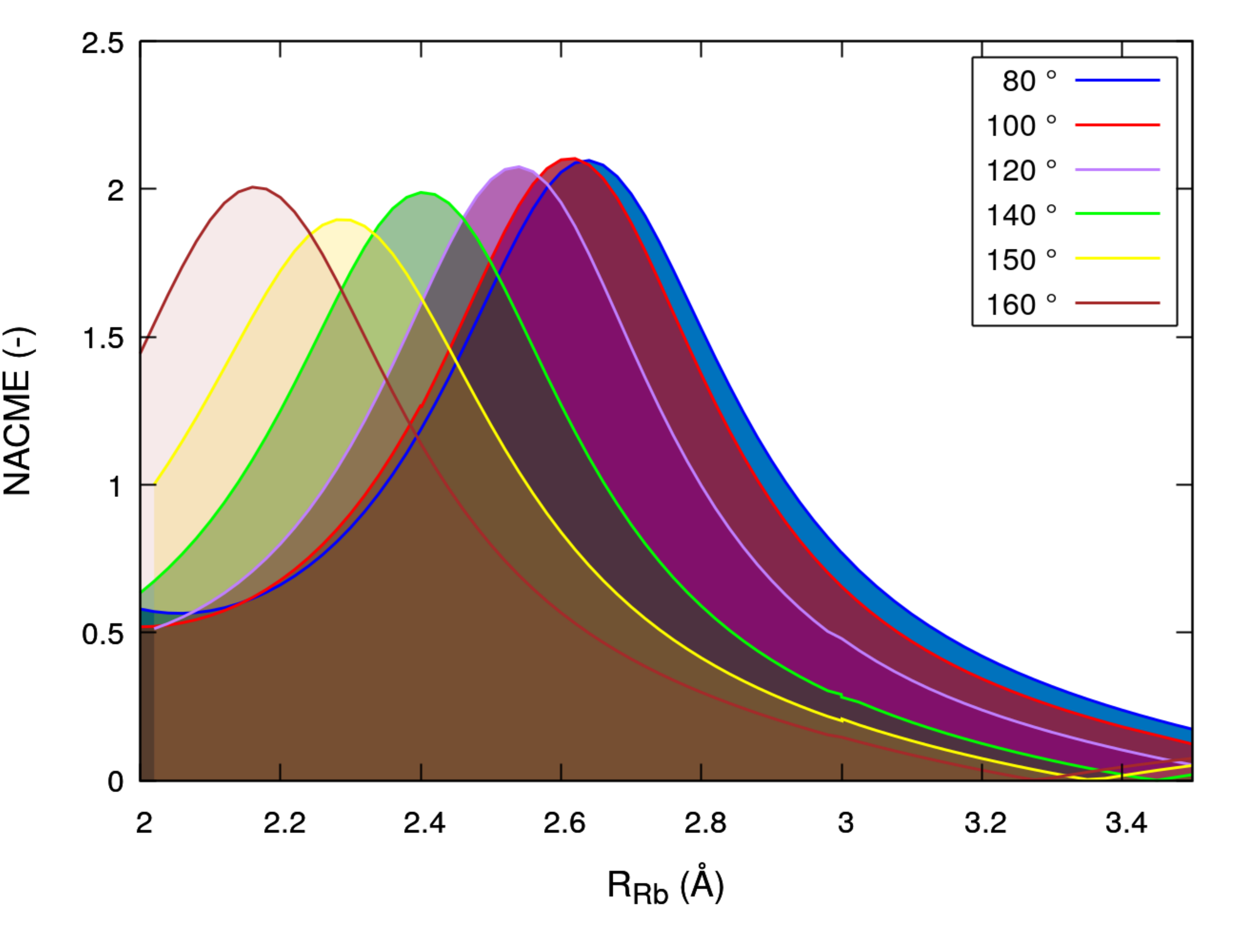}
\caption{Non-adiabatic coupling strength between the X$\text{E}_{1/2}$ and 2$\text{E}_{1/2}$ states of the Rb-OH$^-$ molecular complex for various $\theta$.}
\label{NACME}
\end{center}
\end{figure}
The coupling exhibit the typical bell-shape behaviour of non-crossing states with constant energy spacing. This coupling case can be described by the Rosen-Zener-Demkov model \cite{Nikitin1999, Desouter-Lecomte1977} for which the coupling potential is modeled by a hyperbolic secant:  
\begin{equation}
    V_{12}(\text{R}_\text{Rb})=v_0 sech(\frac{\text{R}_\text{Rb}-\text{R}_0}{\beta})
\end{equation}
and the probability transition between adiabatic states (in atomic units) is given by
\begin{equation}
P_{12}=sin^{2}(\pi v_0 \beta)sech^{2}(\frac{\pi \Delta E \beta}{2\sqrt{2E_k/ \mu}})
\end{equation}
where $\Delta E$ is the energy difference between both non-crossing adiabatic states, $E_{k}$ is the kinetic energy. The parameters $\beta$, R$_{0}$ and $v_0$ were extracted by fitting our NACME calculation. The fitted values are $v_0=1.95$, R$_{0}=4.79$ and $\beta=0.63$. With $\Delta E \approx 7\times10^{-3}, E_k\approx 0.056$ we found the non-adiabatic probability transition $P_{12} \approx 0.15 \%$ for the relevant collision energies. This small probability is mainly due to the large reduced mass of the system and the larger energy gap between adiabatic states in comparison with the avoided crossing case. This justifies the neglect of the electronic to kinetic energy transfer channel in the dynamics.

\subsection{Charge transfer channel}
The probability for an adiabatic passage at the various avoided crossings (at R$_\text{Rb}$ $\approx 8\;\si{\angstrom}$, $7\;\si{\angstrom}$ and $4 \;\si{\angstrom}$ in Supplementary Figure \ref{PES_SO}, left panel) is estimated using the Landau-Zener formula:
\begin{equation}
    P_{12} = exp\big(\frac{\pi (\Delta E)^2}{2\Delta F \sqrt(2E_k/\mu)} \big)
\end{equation}
where $\Delta E$ and $E_k$ are the energy gap and kinetic energy at the avoided crossing point respectively, $\mu$ is the reduced mass and $\Delta F$ is the difference in the slope between the two diabatic curves. We found the diabatic transition probability $P_{12}$ to be close to one for the relevant collision energies. This is primarily due to the small energy gap between the adiabatic states. Typical value of $P_{12}$ are found to be around $0.95$.

The probability to exit via the charge transfer channel Rb$^{-}$+OH is given by $P_{CT}=1-P_{NR}$, where $P_{CT}$ is obtained by summing the probabilities associated to different paths that lead to the non-reactive channel. Owing to the highly diabatic nature of the PES ($i.e$ small Landau-Zener adiabatic transition probability, see above), $P_{CT}$ is very small, around $1.5\%$ for the relevant collision energies. Since the loss from the electronic to kinetic energy transfer channel is also found to be small, the AED reaction is predicted to be the dominant loss channel. 


        %
        %
        %
\renewcommand{\refname}{Supplementary References}        
\bibliographystyle{naturemag}
\bibliography{AD}